# Losses in Plasmonics: from mitigating energy dissipation to embracing loss-enabled functionalities

SVETLANA V. BORISKINA,* THOMAS COOPER, LINGPING ZENG, GEORGE NI, JONATHAN K. TONG, YOICHIRO TSURIMAKI, YI HUANG, LAUREEN MEROUEH, GERALD MAHAN, GANG CHEN

*Department of Mechanical Engineering, Massachusetts Institute of Technology, Cambridge, MA, 02139, USA*
*Corresponding author: sborisk@mit.edu*

Unlike conventional optics, plasmonics enables unrivalled concentration of optical energy well beyond the diffraction limit of light. However, a significant part of this energy is dissipated as heat. Plasmonic losses present a major hurdle in the development of plasmonic devices and circuits that can compete with other mature technologies. Until recently, they have largely kept the use of plasmonics to a few niche areas where loss is not a key factor, such as surface enhanced Raman scattering and biochemical sensing. Here, we discuss the origin of plasmonic losses and various approaches to either minimize or mitigate them based on understanding of fundamental processes underlying surface plasmon modes excitation and decay. Along with the ongoing effort to find and synthesize better plasmonic materials, optical designs that modify the optical powerflow through plasmonic nanostructures can help in reducing both radiative damping and dissipative losses of surface plasmons. Another strategy relies on the development of hybrid photonic-plasmonic devices by coupling plasmonic nanostructures to resonant optical elements. Hybrid integration not only helps to reduce dissipative losses and radiative damping of surface plasmons, but also makes possible passive radiative cooling of nano-devices. Finally, we review emerging applications of thermoplasmonics that *leverage* Ohmic losses to achieve new enhanced functionalities. The most successful commercialized example of a loss-enabled novel application of plasmonics is heat-assisted magnetic recording. Other promising technological directions include thermal emission manipulation, cancer therapy, nanofabrication, nano-manipulation, plasmon-enabled material spectroscopy and thermo-catalysis, and solar water treatment.

## 1. INTRODUCTION

Nanostructures made of materials with high density of free charge carriers can concentrate the incoming light flux to volumes much smaller than the diffraction limit of light. The physical mechanism behind such striking functionality is the excitation of **surface plasmon polaritons** (SPPs) – partially coherent oscillations of free electrons in the material resonantly driven by the external electromagnetic waves [1]. Excitation of SPP modes enables strong intensity enhancement of local electric fields in the vicinity of the nanostructures. The sub-wavelength regions of SPP-enhanced electric field intensity – often referred to as **hot spots** – find use in high-sensitivity optical sensing and spectroscopy as well as in high-resolution imaging [2–8]. Tight optical confinement and hybrid electron-photon nature of SPP modes also offers opportunities for miniaturization of optical networks, their hybridization with electronic circuits, and for designing functional optical metamaterials for the infrared and visible spectral ranges [9,10]. The above advantages of plasmonics helped to fuel high expectations for its rapid expansion into the domain of dielectric- or semiconductor-based optics and photonics technologies [11]. Proposed plasmonic analogs of optical communications components encompassed polarizers [12], modulators [13,14], photodetectors [15], and other functional circuit elements [16]. However, as the amount of research data accumulated, high expectations for the emerging plasmonics revolution in communications were dampened by the hard reality of fast decay and energy dissipation of SPP modes [17,18].

Surface plasmons undergo dephasing and decay on the femtosecond timescale, both by radiative damping and via the formation of energetic or 'hot' charge carriers [19]. Then, the generated hot carriers relax by locally heating the nanostructure and its surroundings within picoseconds. Fast SPP dephasing translates into broad resonant features in their frequency spectra, which are detrimental for sensing applications. It also reduces the signal propagation distances in plasmonic waveguides, leads to the distortion of ultra-fast pulses [20], and limits the energy that can be accumulated by the SPP mode. In turn, the limited energy of an SPP mode caps the maximum local electric field intensity that can be reached. Finally, SPP energy dissipation via absorption – driven by high material losses in metals and other plasmonic materials – results in localized heating of plasmonic elements and nano-circuits [21].

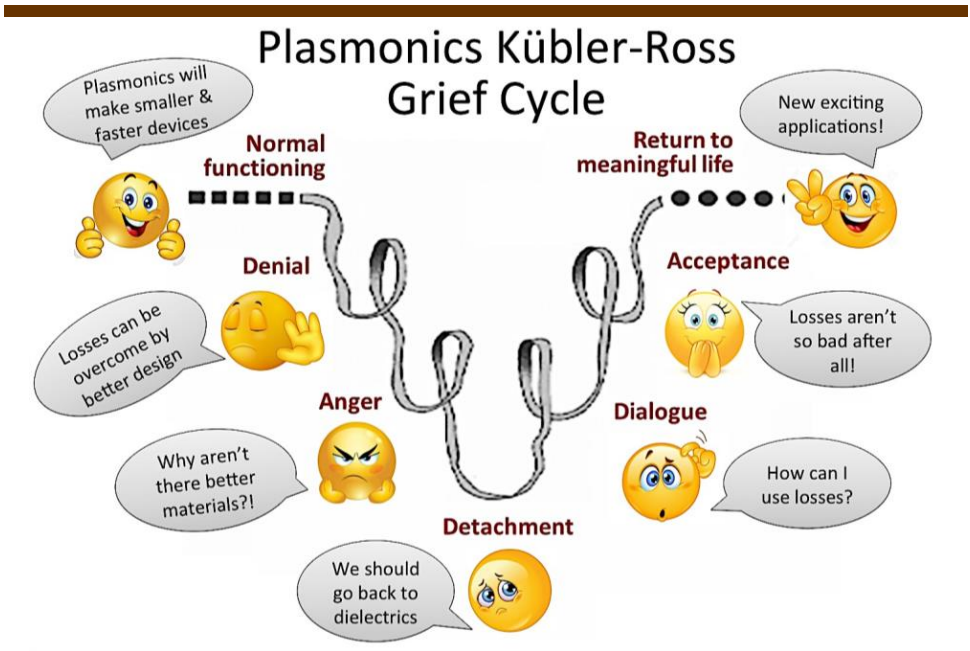

**Fig. 1. Plasmonics Kubler-Ross grief cycle**: evolution of the attitude within the research community towards the role of dissipative and radiative losses in the success of the plasmonics research enterprise.

Confronting this reality spurred many new directions in the plasmonics and nanophotonics research, some aimed at either reducing or compensating the losses, and others offering alternative applications by leveraging them (see Fig. 1). In this paper, we review several approaches to reduce or mitigate both dissipative and radiative losses in plasmonics, with the focus on discrete nanostructures supporting localized SPP modes rather than on surface plasmon polariton waveguides. In particular, we show how sensing applications – where plasmonics already plays a major role – significantly benefit from improvements in spectral selectivity and local field intensity achieved by suppressing both loss mechanisms. We also discuss a myriad of emerging applications of plasmonics that stem from harnessing rather than fighting material dissipative losses.

## 2. PLASMON WAVES AND THE ORIGIN OF PLASMONIC LOSSES

### 2.1 Electron plasmas and their collective oscillations

As the name suggests, plasmonics deals with generation, detection, control, and use of plasmons as the information and energy carriers. Unlike photons, which are quanta of transverse electromagnetic waves carrying energy $\hbar\omega$, plasmons are quanta of collective longitudinal charge-density oscillations in plasmas with energy $\hbar\omega_p$ (for bulk plasmons). Here, $\omega$ is the angular frequency, $\hbar$ is the reduced Plank's constant, $\varepsilon_0$ is the vacuum permittivity, and the plasma frequency $\omega_p = (n_e e^2 / m_e \varepsilon_0)^{1/2}$ characterizes the speed with which electrons with electric charge $e$, effective mass $m_e$ and density $n_e$ *collectively* react to an applied external electric potential. Plasmas are media with large densities of free charge carriers. Valence band electrons in metals or doped semiconductors form electron plasmas, with electron densities $n_e \sim 10^{22}$ cm$^{-3}$ and $\sim 10^{19}$ cm$^{-3}$, respectively. For metals, which have high density of free electrons, plasma frequency typically is on the order of $10^{15} Hz$. The interaction between plasmons and photons on the interfaces between plasmonic and dielectric materials creates another quasi-particle, called a surface plasmon polariton. Although photons, plasmons, and surface plasmon polaritons are quantum mechanical bosonic oscillations, they can often be treated quasi-classically.

The energy-momentum dispersion relation for a propagating photon wave in a vacuum has a familiar form shown in Fig. 2a, $|\mathbf{k}|^2 = k^2 = \omega^2/c^2$, where $c$ is the speed of light in vacuum and $\mathbf{k}$ is the wavevector. Optics and photonics typically use dielectrics and semiconductors as host media for photon propagation. In their transparency bands, dielectrics resist electromagnetic waves by reducing the wave phase velocity by a factor $\sqrt{\varepsilon}$, where $\varepsilon = \varepsilon(\omega)$ is the dielectric permittivity of the medium. In turn, the permittivity of plasmonic materials should capture (i) the response of the charge carriers to both an external electric potential and the internal potential imposed by the material crystal structure, (ii) the screening of the Coulomb potential by the interacting charges, and (iii) the decay of collective oscillations due to collisionless single-particle excitations (i.e. Landau damping) and collision damping processes.

The permittivity of plasmonic materials can be semi-classically approximated by using a damped harmonic oscillator model $\varepsilon(\omega) = 1 - \omega_p^2 / (\omega^2 + i\gamma\omega - \beta k^2)$, and is known as the **Drude permittivity** [22]. As we will see in the following, in this modern form applicable to modeling metal quantum plasmas, it should perhaps be accurately referred to as the Drude-Sommerfeld-Bloch-Landau permittivity. An empirical damping parameter $\gamma$ ($\gamma < \omega_p$) defines the lifetime $\tau = 1/\gamma$ of collective plasma oscillations due to energy dissipation through various decay channels. Non-local coefficient $\beta$ describes the Landau damping process. It depends on the magnitude and direction of the average velocity of electrons that participate in the collective motion forming the plasmon wave. Non-local behavior of the permittivity only becomes pronounced at high values of the SPP wavevector, and is typically ignored in situations when low-*k* SPP modes are excited. While the Drude expression above is limited to describing the free-electron contribution to the permittivity of a plasmonic material, it can be further generalized to include the effect of the inter-band electron transitions (section 4).

### 2.2 Energy-momentum dispersion of surface plasmon polariton modes

At material interfaces, bulk plasmons can couple with photons to form hybrid surface electromagnetic modes – surface plasmon polaritons [23,24]. The energy-momentum dispersion of SPP modes strongly depends on the shape of the interface on which they are excited. In the simplest case of a planar interface between a vacuum and a plasmonic material, it takes the following familiar form: $k_\parallel(\omega) = (\omega/c) \cdot (\varepsilon/(\varepsilon+1))^{1/2}$ [25,26]. This dispersion is shown in Fig. 2a, where the real wavevector component parallel to the interface is plotted as a function of frequency. In the case of the local damping-free approximation of the metal permittivity (i.e., $\beta = 0$, $\gamma = 0$), SPPs propagate along a planar metal-vacuum interface with frequencies ranging from zero at $k_\parallel = 0$ and approaching the asymptotic value $\omega_{SPP} = \omega_p/\sqrt{2}$ at $k_\parallel \to \infty$ (lower blue branch in Fig. 2a). In this frequency range, the real part of material permittivity becomes negative, making possible the formation of the surface polariton mode. The upper blue branch in the dispersion corresponds to the bulk plasmon oscillations.

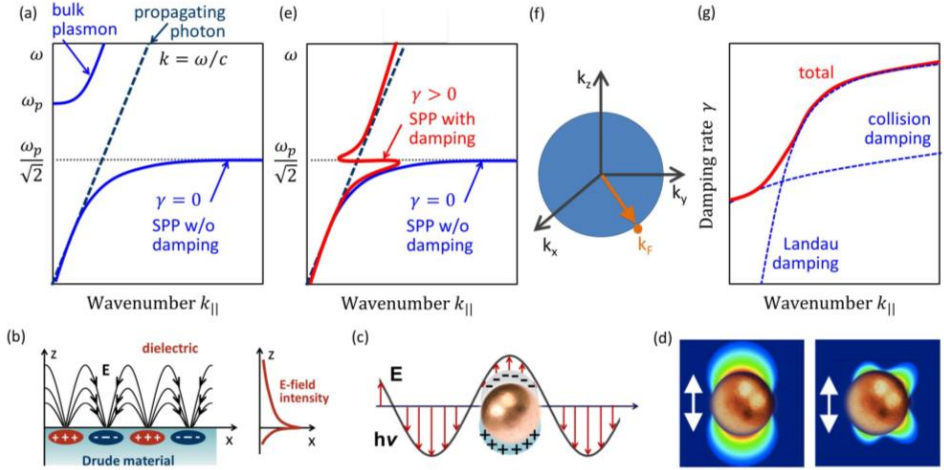

**Fig. 2. Photon, plasmon and surface plasmon polariton dispersion in the energy-momentum space.** (a) Dispersion relations for propagating photons (also known as the light line; dashed navy line), bulk plasmons (upper blue branch), and SPPs on a planar metal-vacuum interface without damping (lower blue branch). (b) Schematic of the charge and electric field distribution in a SPP mode on a planar interface. (c) Schematic of a localized SPP mode excitation on a plasmonic nanosphere. (d) Electric field intensity distributions around a nanosphere excited at the frequencies of its dipole and quadrupole SPP modes, respectively. (e) Comparison of the dispersion relations for SPPs on a planar interface with (red line) and without (blue line) damping. (f) Spherical Fermi surface in the momentum space of free electrons in metal within the Drude-Sommerfeld model. The electron states within the sphere are filled, while the higher-energy higher-momenta states outside the sphere are empty. The electrons on the surface are characterized by energy $E_F$ (Fermi energy) and momentum $k_F$ (Fermi momentum). (g) Collision and Landau damping rates scaling with the increasing SPP momentum.

The lower SPP branch is always appearing to the right of the light line (i.e., of the dispersion curve for propagating photons, dashed navy line in Fig. 2a). This is a manifestation of the fact that surface plasmon polaritons have larger momenta (and shorter wavelengths) than plane electromagnetic waves at the same frequency. Hence, SPPs on ideal metal-dielectric interfaces are nonradiative in nature, as the momentum conservation law prohibits their decay via propagating photon emission into the far field. The surface nature of the planar SPP mode is schematically illustrated in Fig. 2b.

SPP mode dispersion becomes quantized in the case of photon and plasmon modes coupling on the surfaces of nanoparticles and finite-size nanostructures [27–29]. The resulting localized SPP modes are characterized by distinct resonant frequencies and different angular momenta. These localized SPP modes can be excited by propagating photons (Fig. 2c), and, by reciprocity, can radiatively decay by emitting photons. The lowest energy and momentum SPP modes are dipole and quadrupole modes, whose local electric intensity profiles are shown in Fig. 2d. Radiative decay of SPPs on planar surfaces is also possible due to surface roughness or curvature imperfections.

Large-momentum states appearing in the SPP dispersion (Fig. 2a) in the absence of damping ($\gamma = 0$) offer promise of achieving high density of photon states (DOS) within a narrow frequency range. This promise made plasmonics an attractive technology for surface-enhanced spectroscopy and optical sensing [2] and fueled high expectations for other applications [1,11]. However, the dispersion of a damped SPP wave (i.e., with $\gamma > 0$, red line in Fig. 2e) reveals the cut-off of high-momentum states, which reduces the SPP DOS and limits the field spatial confinement. Unfortunately, excitation of high-momentum SPP states is unattainable due to a combination of intrinsic mechanisms associated not only with the material imperfections but also with the very nature of SPP oscillations [17,18,28].

### 2.3 The origin of plasmonic losses: collision vs collisionless SPP energy dissipation

First of all, due to the quantum nature of electrons, the largest wave vector of plasmon oscillations cannot exceed the **Fermi wave vector** ($k_F \sim 1.2 \cdot 10^8 \, cm^{-1}$ for noble metals), known as the 'quantum limit'. This limit stems from the fact (originally pointed out by Sommerfeld) that the free charge carriers in plasmonic materials

> High-momentum SPP states are unattainable not only due to the material losses but also because of the very nature of plasmon oscillations. This imposes limits on the achievable confinement and DOS of surface plasmon modes.

should be treated as quantum-mechanical particles (fermions), which obey the Fermi-Dirac statistics. Accordingly, inside metals, the electrons in the ground state occupy the energy levels up to **Fermi energy** $E_F = (\hbar^2/2m_e) \cdot (3\pi^2 n_e)^{2/3}$, and their wave vectors are restricted by the Fermi wave vector $k_F = (3\pi^2 n_e)^{1/3}$ [30]. The surface in the reciprocal (i.e., momentum) space that separates occupied and unoccupied electron states at zero temperature is known as the **Fermi surface**. In the Sommerfeld free-electron model the Fermi surface is a sphere (Fig. 2f). It serves as a reasonable approximation for noble metals [31]. However, it can have complicated shapes in other materials, which are defined by the material crystal lattice and can be calculated by using the Bloch theory [32]. The high-energy electrons on the spherical Fermi surface have velocity $v_F = (2E_F/m_e)^{1/2}$, known as the Fermi velocity.

Plasmon and surface plasmon polariton oscillations are damped by elastic and inelastic scattering of electrons on other electrons, on lattice vibrations (phonons), and on lattice defects. Scattering-related damping parameter – often referred to as bulk damping $\gamma_b$ – is an average collision frequency resulting from various internal scattering processes: $\gamma_b = \gamma_{e-e} + \gamma_{e-phonon} + \gamma_{defect} + \ldots$. The typical damping rate due to electron-phonon scattering in metals is high ($\gamma_{e-phonon} \sim 10^{14} \, Hz$), driven by the large density of electron states above the Fermi level available for the electron transition in the absence of the electron bandgap. It grows with temperature, while the scattering by material defects provides a constant term to the damping parameter, which does not disappear even at zero temperature. Electron-electron scattering rate increases with frequency as $\gamma_{e-e} \sim 10^{15} \cdot (\hbar\omega/E_F)^2 \, Hz$, and becomes comparable with $\gamma_{e-phonon}$ in the visible frequency range [18,28].

Bulk damping rate $\gamma_b$ relates to the mean free path (mfp) $\mathsf{L}_e$ of electrons in the material as $\gamma_b = v_F/\Lambda_e$, where $v_F$ is the Fermi velocity. Accordingly, SPP oscillations on nanoparticles with sizes comparable to or smaller than the electron mfp experience additional damping due to electron collisions with the particle surface, $\gamma = \gamma_b + \gamma_{sc}$. The surface collision damping rate is inversely proportional to the average distance $a$ an electron can travel before scattering off a surface, $\gamma_{sc} = v_F/a$. The Fermi velocity and damping constant for bulk Au are $v_F = 1.4 \cdot 10^6 \, ms^{-1}$ and $\gamma_b = 15 \, fs$ [33] at room temperature, which translates to an electron mfp of approximately 20 nm. The electron mfp in other metals are also on the order of tens of nanometers, and thus surface collisions play a role in the SPP damping on nanostructures of comparable size [28]. Surface roughness also contributes to the surface collision damping of SPP modes, since fabrication of ultra-smooth surfaces is challenging [34]. Surface oxidation of plasmonic materials, on the other hand, can play both positive and negative role in their functional performance [35,36].

Finally, even in the absence of electron collisions with phonons, defects, material interfaces, and with other electrons, plasmon oscillations that are spatially localized into a region comparable to the Debye length *in any dimension* are heavily damped. **Debye length** is the distance traveled by an average electron during a plasmon wave oscillation period, which is $\lambda_D = v_F/\omega_p$ for quantum plasmas in metals. This phenomenon is known as **Landau damping**, and arises when electrons with the same velocity as the plasmon phase velocity are accelerated by the wave and extract energy from it, damping plasmon oscillations in the process [37–40]. Landau damping effectively represents a plasmon-electron scattering process, in which a plasmon wave loses a single quantum to generate an electron-hole pair. Plasmon decay rate associated with Landau damping scales with the SPP confinement length (and thus with the SPP wavevector) as $\gamma_L \sim \omega_p \cdot (k_D/k)^3 \cdot \exp(-k_D^2/2k^2)$, where $k_D = \omega_p/v_F$ is the Debye wavevector. Landau damping rate exceeds collision-induced damping rates for tightly-confined high-momentum plasmon modes (Fig. 2g) [41,42]. For example, while the Debye length for gold electron plasma is about 6Å, surface plasmons on gold surfaces experience strong confinement-induced damping when confined to a spatial region $a_c \sim 10 nm$ [43]. It should be noted that Landau damping is also the cause of the non-local behavior of the real part of the longitudinal permittivity [37], which only becomes pronounced in the regime of tight field confinement and affects high-momentum SPP modes. The non-local parameter $\beta$ in the Drude permittivity function scales with the electron velocity and contributes differently to the longitudinal and transverse permittivities. In the random-phase approximation of quantum plasmas in metals $\beta_\parallel = (3/5)v_F^2$ and $\beta_\perp = (1/5)v_F^2$ [44].

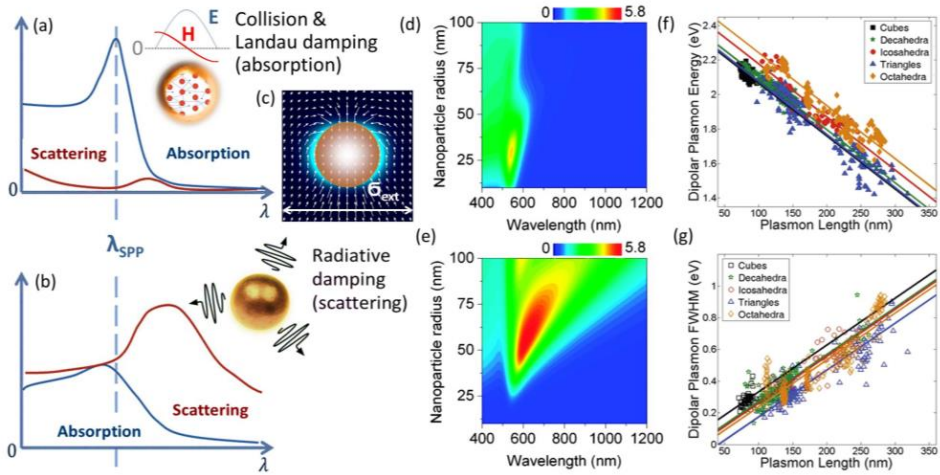

**Fig. 3. Surface plasmon dephasing and decay via dissipative and radiative damping mechanisms.** (a,b) Large resonant absorption dominates surface plasmon decay on the nanostructures size-tuned to exhibit a dipole resonance at a wavelength close to the plasma frequency in the material. In this case, electric field intensity significantly exceeds magnetic field intensity, resulting in the energy being stored and dissipated by the kinetic motion of charge carriers for a half-cycle of the oscillation. (c) Extinction cross section and the optical energy flow around a nanosphere at the dipole SPP resonance frequency. (d,e) Absorption (d) and scattering (e) efficiencies of a gold nanosphere in water as a function of the excitation wavelength and the particle radius. Larger particles with the dipole SPP resonance red-shifted away from the plasma frequency and the material direct absorption band dissipate energy via radiative damping. Panels d,e are reproduced with permission from [45], ©ACS. (f,g) The energy stored in the SPP dipole mode decreases and the resonance linewidth broadens due to increased radiative damping

of surface plasmons on larger particles regardless of the particle shape (reproduced with permission from [46],©ACS).

Another intuitive way to understand Landau damping in the regime of the strong electromagnetic field confinement is to recall that SPP waves are hybrid photon-electron states [12,31]. During the oscillations of a purely electromagnetic mode, every half period the energy is transferred back and forth between the electric field energy and the magnetic field energy. In the case of tight spatial confinement, local magnetic field becomes a tiny fraction of the local electric field (Fig. 3a). The SPP oscillations can still be sustained because the hybrid wave energy is stored not only in the electromagnetic field but also in the form of kinetic energy of electrons. This however causes the SPP wave – which involves collective oscillations of *ground-state electrons* – to lose energy due to excitation of energetic electron-hole pairs.

## 2.4 The effect of plasmonic losses on SPP modes spectra and local field enhancement

In practical implementation of SPP oscillations, the SPP damping rate translates into the broadened linewidth of the SPP resonance excited e.g., by the external source of propagating photons (Figs. 3a,b). The SPP linewidth is broadened by both dissipative and radiative damping mechanisms. Confinement-induced SPP dephasing and damping becomes especially pronounced in the case of small plasmonic nanoparticles, when both Landau damping and interface-collision damping cause the SPP extinction via absorption (Fig. 3a). For larger particles, radiative damping of the SPP mode via photon scattering becomes more dominant than internal damping mechanisms (Fig. 3b). Radiative damping of SPP modes on nanoparticles with sizes comparable to the excitation wavelength is often phenomenologically accounted for via a volume-dependent radiative damping rate $\gamma_r = C_r \cdot V$, where is $C_r$ the constant that characterizes the damping efficiency and $V$ is the nanoparticle volume [18,47].

Direct photon absorption (i.e., not occurring via SPP excitation; sections 4,5) plays a role in photon energy dissipation by both, small and large metal nanostructures under excitation by broadband optical sources (compare Figs. 3a,b). The ability of the plasmonic nanostructure to absorb and scatter incoming photons is typically characterized by their absorption ($\sigma_{abs}$) and scattering ($\sigma_{sc}$) cross-sections, and their sum comprises the **extinction cross section** $\sigma_{ext} = \sigma_{abs} + \sigma_{sc}$ [28,29]. These cross-sections quantify by how much the light extinction by an object in the photon path exceeds the light extinction due to blocking the area of the same geometrical cross-section. At the frequencies corresponding to the excitation of SPP resonances, these cross sections may significantly exceed their geometrical counterparts [29,48], indicating that plasmonic nanoparticles can scatter and absorb more than the light energy propagating through the area of their geometrical cross-section. The physical mechanism underlying such enhanced scattering and absorption is illustrated in Fig. 3c, which shows bending of the Poynting vector in the area around a metal nanosphere excited at its dipole SPP mode [48,49]. As a result, the energy is being recycled through the volume of the nanosphere and dissipates via absorption and scattering loss channels.

Figures 3d and 3e quantify the absorption and scattering efficiencies (i.e., ratios of the corresponding cross-sections to the geometrical cross-section) for an Au sphere immersed in a dielectric medium with refractive index $n = \sqrt{\varepsilon} = 1.44$ [45]. Both efficiencies are plotted as a function of the nanosphere size and the excitation wavelength, and clearly demonstrate high absorption in smaller nanoparticles and enhanced scattering induced by larger ones. Note that scattering-induced radiative damping of the SPP mode results in significant broadening of the SPP resonance (Fig. 3e), which, despite low absorption losses, reduces the amount of energy that can be stored in the SPP mode. This in turn reduces the maximum local electric field intensity that can be achieved on the nanoparticle surface [38]. Local field intensity can be somewhat re-distributed on the particle surface via a so-called 'lightning rod effect' by introducing sharp edges, protrusions or indentations [50,51]. However, the total amount of energy stored in the SPP mode still decreases with the increase of the particle size [52]. This is illustrated in Fig. 3f for a number of nanoparticle shapes that feature sharp corners. Figure 3g further shows that the observed drop in the stored energy is driven by the increased mode linewidth, which indicates faster mode dephasing due to radiative damping [52].

## 3. SCULPTING PLASMONIC SPECTRA VIA DESTRUCTIVE INTERFERENCE

The limit imposed by plasmonic losses on the amount of energy stored in the SPP mode reduces the efficiency of plasmonic devices that make use of the strong local field enhancement. These include surface-enhanced Raman scattering (SERS) [2,5,45,53], surface-enhanced fluorescence (SEF) sensing [7,54] surface-enhanced infrared absorption (SEIRA) spectroscopy [55–57], and optical nanoparticle detection [58] among others. For example, SERS spectroscopy makes use of excitation of SPP modes to enhance the inelastic (i.e. Raman) scattering of photons by phonons in target materials, which can thus be fingerprinted by recording the spectra of Raman-shifted photons. The standard **figure of merit (FOM)** of a SERS sensor is often defined as the product of the local electric field intensity enhancement at the pump wavelength $\omega_0$ and the corresponding enhancement at either Stokes- or anti-Stokes signal wavelength, $\omega_s$: $FOM_{SERS} = |E(\omega_0)|^2 / |E_0(\omega_0)|^2 \cdot |E(\omega_s)|^2 / |E_0(\omega_s)|^2$, where $E_0(\omega)$ is the electric field in the absence of the plasmonic nanostructure. In turn, SPP resonance broadening induced by either scattering or absorption losses further degrades efficiency of plasmonic sensors. The standard figure of merit of a plasmonic sensor that detects a shift of the resonance wavelength $\lambda_{res}$ due to the changes in its environment is $FOM_{sen} = sensitivity/linewidth$ [59,60]. Sensor sensitivity $S = \Delta\lambda_{res}/\Delta n$ is defined as the resonance shift resulting from the environmental changes (e.g., from the ambient refractive index change), and scales with the local field enhancement [58].

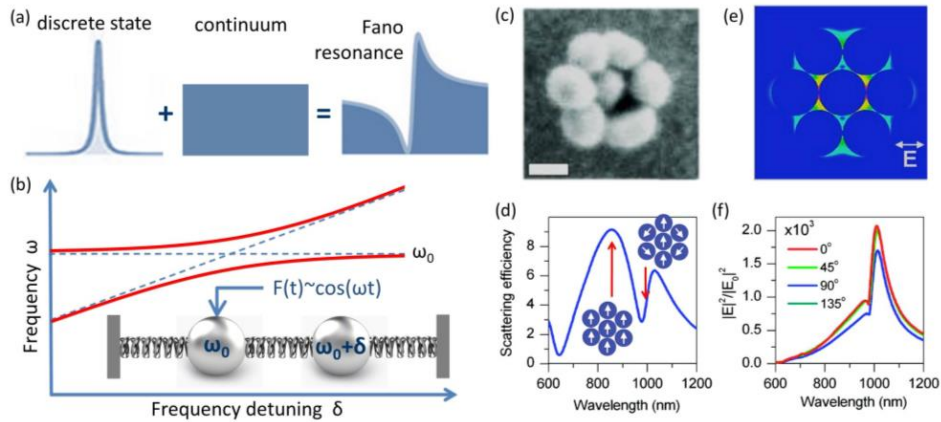

**Fig. 4. Fano interference yields narrower-linewidth resonant features in the optical spectra of plasmonic nanostructures.** (a) Schematic of the Fano resonance resulting from coupling of a discrete state with a Lorentzian spectral shape and a continuum of states. (b) A system of coupled oscillators that provides a classical example of the Fano interference mechanism. Two masses that oscillate with mismatched natural frequencies $\omega_0$ and $\omega_0 + \delta$ are coupled via springs and are driven by an external periodic force with frequency $\omega$. Hybridized (bonding and anti-bonding) modes of the coupled oscillator system (solid red lines) are plotted as a function of the resonant frequency detuning $\delta$. (c) A heptamer plasmonic nanocluster composed of seven Au nanospheres. (d) Scattering spectrum of the heptamer exhibits a Fano-shaped spectral feature formed by the interference between SPP modes of the constituent nanospheres. (e) Local electric field intensity distribution inside the nanocluster. (f) Maximum electric field intensity on the heptamer as a function of wavelength. Panels c-f are modified with permission from [45], ©ACS.

However, the linewidth of a resonant feature in the optical spectrum of a plasmonic sensor is a parameter that can be improved via the sensor configuration engineering. Accordingly, a lot of effort has been invested into shaping optical spectra of plasmonic structures to achieve spectral features with narrow linewidths. Although in many cases the SPP mode damping is not actually reduced, this approach has been very successful in designing structures with narrow spectral features, which can be extremely beneficial for decreasing the sensor detection limit. One approach to sculpting optical spectra of plasmonic nanosensors is via a combination of constructive and destructive interference of SPP resonances. In particular, Fano-type interference has been extensively used to reduce the resonance linewidths [61,62]. First described by Ugo Fano in the case of inelastic scattering of electrons [63], **Fano resonance** is often explained as a result of the constructive and destructive interference of a narrow Lorentzian resonance with a continuum of states (Fig. 4a) [61]. This interference produces a spectral feature with an asymmetric lineshape (see Fig. 4a) characterized by the following functional form: $I(\omega) \sim (A\gamma_f + \omega - \omega_f)^2 / ((\omega - \omega_f)^2 + \gamma_f^2)$, where $A$ is the asymmetry parameter, and $\omega_f$ and $\gamma_f$ represent the position and the linewidth of the Fano spectral feature.

Another simple model of the Fano interference is a driven classical coupled oscillator (Fig. 4b), where multiple energy-transfer pathways between the oscillator and the external source interfere constructively and destructively [61,64]. One of the hybridized modes of the coupled-oscillator system resulting from such interference (red lines in Fig. 4b) exhibits an asymmetric Fano lineshape. In the case of a strong damping of one of the oscillators, the coupled system frequency spectrum closely resembles the Fano shape resulting from a discrete mode interference with a continuum (Fig. 4a). In close analogy with coupled classical oscillators, the interference either between SPP modes with different angular momenta supported by a single plasmonic nanoparticle or between different SPP modes in coupled-particle structures can be used to generate asymmetric Fano spectral features [45,61,62,65–69].

Finite-size clusters (so-called oligomers) of plasmonic nanoparticles have been shown to support Fano resonances arising from the multi-particle interference. One well-studied example of a nanoparticle cluster exhibiting a Fano lineshape is a heptamer cluster composed of a central particle surrounded by six other nanoparticles (Fig. 4c) [45,62]. The heptamer bonding and anti-bonding hybridized modes result from the in-phase and out-of-phase coupling between SPP dipole modes of central particles and particles on the sides of the outer ring (see insets to Fig. 4d). The in-phase bonding mode is broadened by strong radiative damping and serves as an analog of mode continuum. The anti-bonding mode punctures this 'continuum' and introduces a Fano resonance in the heptamer optical spectrum (Fig. 4d). The Fano feature exhibited by a cluster is significantly narrower than the dipole SPP mode of a single particle of equivalent size.

Narrower linewidths translate into higher FOM values observed in nanocluster-based SPP sensors (up to ~11) [59,70] than those predicted and measured in single-particle SPP sensors (from ~1 to ~5.5) [60,70]. However, Fano resonances of plasmonic nanoclusters are still very broad in comparison to resonant modes of dielectric and semiconductor clusters [71,72]. Furthermore, the Fano resonance in a heptamer cluster does not yield local electric field enhancement. Maximum local field intensity is comparable to that in a dimer or a triangular cluster of the same particles separated by the same distances [45].

However, destructive interference between SPP modes of plasmonic nanoparticles arranged into cluster-like configurations can be exploited to achieve narrow Fano-line spectral features that are also accompanied by a very strong field enhancement. One of such structures can be composed of five nanoparticle dimers arranged into a short linear chain [49,73,74] (Fig. 5a). This nanostructure exhibits a Fano-like resonance in its optical spectrum (Fig. 5b) and strong local electric field intensity enhancement in the dimer gaps (Fig. 5c and Fig. 5d, red solid line) under plane wave illumination polarized along the dimers axes [49,73,74]. The strong field enhancement is achieved due to optical energy recycling through localized optical vortices (Fig. 5e), which form around four points of destructive interference in the vicinity of the nanostructure. The positions of the points of destructive interference (i.e., points of complete darkness) are visible as local dark spots in Fig. 5e, which shows the Poynting vector rotation in the plane cutting through the center of the dimer gaps perpendicular to the nanostructure. At the points of zero field intensity, optical phase becomes undefined, which forces the optical powerflow to twist around resulting phase singularities [75–77]. Overall, phenomena of Fano resonance formation and nanoscale vortical powerflow are intrinsically connected [78].

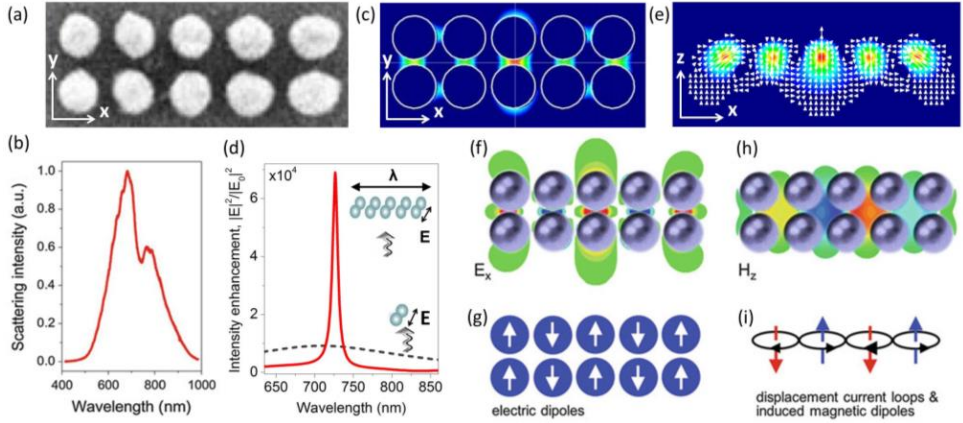

**Fig. 5. Destructive interference drives local powerflow in plasmonic nanostructures and can reduce both dissipative and radiative losses.** (a) Scanning electron microscopy (SEM) image of a linear dimer chain of gold nanoparticles separated by nm-scale gaps. (b) An experimental scattering spectrum of the dimer chain that features a Fano-like narrow resonance. (c) Calculated local electric field intensity enhancement in the dimer gaps. (d) Calculated electric field intensity enhancement in the central dimer gap of the structure as a function of wavelength (red solid line). The gray dotted line shows the intensity enhancement in a gap of a single standalone dimer. (e) The Poynting vector distribution in the plane perpendicular to the dimer axes cutting through the center of the dimer gaps. The plane wave is incident from the bottom and its energy circulates within the nanostructure by twisting around the points of destructive interference (i.e., phase singularities). (f) Local electric field distribution within the nanostructure. (g) Orientation of the dipole SPP modes of individual nanoparticles. (h) Local magnetic field distribution within the nanostructure. (i) Displacement current loops and magnetic dipoles induced by the current circulation. Panels b,d,f-i are adapted with permission from [74] ©Springer.

Because in Fig. 5 the optical power is being recycled through the dimer gaps rather than through the nanoparticle volumes as in the case of a single nanoparticle (Fig. 3c) or a heptamer cluster (Figs. 4c-f), absorption losses are reduced. Radiative damping is also reduced owing to photon recycling inside a mirrorless resonator formed by the linear cluster (Fig. 5b). As a result, electromagnetic energy can accumulate in the hybridized SPP mode yielding strong local intensity enhancement shown in Fig. 5d. Furthermore, the narrow linewidths of the resonant features in linear nanoparticle dimer chains translate into high FOM values of SPP sensors (e.g., FOM=85 in [74]).

It should also be noted that the electromagnetic energy accumulates because a significant portion of it is stored in the form of the magnetic field rather than in the form of the kinetic energy of electrons as in the case of standalone SPP nanoparticles. As shown in Figs. 5f-5i, the circulating displacement currents generated in the nanostructure at the resonant frequency induce magnetic moments. This effect is completely analogous to the generation of time-varying magnetic fields by conducting currents circulating through wire loops. In particular, antiphase magnetic moments form in the adjacent loops of displacement current, and their mutual coupling along the chain gives rise to an artificial "antiferromagnetic" behavior (Fig. 5i). Excitation of magnetic moments that leads to the increase of the local magnetic field has also been observed in heptamer plasmonic molecules at the frequency of their anti-bonding hybridized SPP mode (see right inset in Fig. 4d) [79]. However, in that case a significant portion of the electromagnetic energy is re-circulated through the metal volume, which increases dissipative losses.

We can conclude that some reduction of dissipative and radiative losses in plasmonic nanostructures can be achieved by sculpting their optical spectra via the mechanism of **destructive interference**. The best results can be achieved if the electromagnetic energy is re-circulated through the nanostructure outside of the metal volume and is stored in the magnetic field rather than in the kinetic energy of electrons.

### 4. NEW PLASMONIC MATERIALS

Despite some progress in engineering narrow-linewidth SPP spectral features in metal nanostructures, high dissipative losses still limit their application range. One of the limitations of the noble metals such as gold and silver stems from the **direct absorption** of high-energy photons from deep and flat $d$-bands in their **electronic bandstructure**. The Fermi surface that separates occupied and unoccupied electron states at zero temperature in the reciprocal (i.e., momentum) space only has a spherical shape within the Sommerfeld free-electron model (Figs. 2f). In real metals, the Fermi surface can be very different from a sphere, ranging from nearly-spherical shapes for noble metals such as silver (Ag) and gold (Au) (Fig. 6a) to complex tubular surfaces like that of lead (Pb) [31,32,80]. Bloch's inclusion of the crystal lattice periodic potential into the free-electron model led to the realization that the electron states are not distributed uniformly within the Fermi sphere but rather fill the distinct electronic bands (Fig. 6b) [81]. To describe the motion of electrons in a given band with a Drude model, the electrons can be treated as free particles that possess an effective mass instead of their real inertial mass.

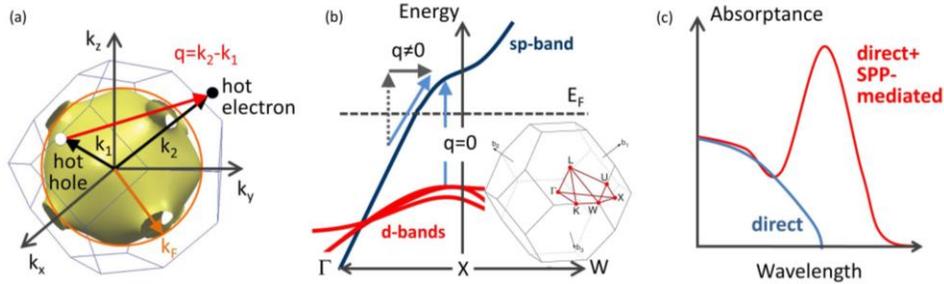

**Fig. 6. Fermi surface, electronic bandstructure, and plasmonic excitations in metals.** (a) Fermi surface of Au (golden surface) and its spherical approximation within the free-electron Drude-Sommerfeld model (orange). The electron transition between different energy states is accompanied by the momentum change. (b) Au electronic bandstructure around the X symmetry point and the Fermi energy level. The inset shows the first Brillouin zone of the Au face-centered cubic lattice and its high symmetry lines and points. The blue arrows show direct ($q=0$) and indirect electronic transitions. (c) Wavelength spectra of metal absorptance due to direct photo-induced electron transitions (blue line) and a combination of direct and surface-plasmon-mediated indirect transitions (red line).

In general, the effective mass is energy-dependent, but a single-valued effective mass serves as a good approximation for electrons at the top of almost-filled bands or at the bottom of almost-empty bands, where the dispersion curves are nearly parabolic. Such electrons with energies within $\sim k_B T$ interval around the Fermi surface typically make the most significant contribution to the charge transport and SPP excitation processes. Electrons from deep fully filled bands below the Fermi level (such as e.g. Au $d$-bands shown in Fig. 6b) cannot be thermally excited at room temperatures and their contribution is not accounted for in the Drude model. However, they can be easily excited via direct photon absorption (Figs. 6b) if the photon energy is larger than the direct energy gap between the $d$-bands and the empty states in $sp$-bands above the Fermi level. The contribution of these **interband transitions** to the complex permittivity of metals can be classically accounted for by including one or more additional terms into Drude permittivity function: $\varepsilon(\omega) = \varepsilon_{Drude} + \sum_{(j)} G_j(\omega)$, where $G_j(\omega)$ is the contribution from the *j*-th interband transition. As the simplest classical approximation, a Lorentz oscillator can be used to describe each transition as a contribution from a bound charge, $G_j(\omega) = A_j / (\omega_{0j}^2 - \omega^2 - i\omega\tilde{\gamma}_j)$, where amplitude $A_j$, natural frequency $\omega_{0j}$ and damping $\tilde{\gamma}_j$ of the bound charge are fitting parameters [82]. More advanced analytical critical-points models, which satisfy the Kramers-Kronig relation between the real and imaginary part of the permittivity have also been developed [83] as well as multi-parametric models that take into account band structure of materials [84]. These more sophisticated models can reproduce most experimentally observed permittivity functions.

Even high-energy photons carry very little momentum and cannot induce indirect excitations of electrons within the *sp*-bands, which require additional momentum *q* (Fig. 6a,b). This extra momentum can be supplied by interaction with phonons (i.e., low-energy yet high-momentum quanta of lattice vibrations), via excitation of two hot electrons with half the photon energy and the opposite momenta, or via excitation of high-momentum SPPs that eventually decay and transfer energy and momentum into the excited electron-hole pairs [43]. The electron transition rates associated with the former two multi-particle excitation processes are low, which reflects the low probability of such events. Due to the presence of the energy bandgap, direct absorption via interband transitions exhibits a clear wavelength threshold as shown in Fig. 6c. In contrast, SPP-mediated indirect excitation of electrons in the *sp*-bands via Landau damping results in very strong resonant absorption of photons with energies well below those that can induce direct electron transitions (Fig. 6c). This opens up opportunities for creating plasmonic absorbers with tailored wavelength spectra for a variety of applications in photon energy harvesting and conversion. In particular, absorption losses can be decreased if the nanostructure is designed to support an SPP resonance outside of the spectral range where direct photon absorption dominates.

The spectral position of the SPP resonant peaks can be tuned not only by the nanostructure design as seen in Figs. 3-5, but also by the choice of plasmonic material. Since the plasma frequency scales with the density of free electrons, materials with lower electron densities than metals (e.g., doped semiconductors) can exhibit SPP resonances at longer wavelengths. Furthermore, *p*-doped semiconductors can also exhibit plasmonic activity. The contribution of holes to the Drude permittivity is of the same sign as for the free electrons, as the material conductivity scales with the second power of the elementary charge as $\omega_{p,j} = (n_j e^2 / m_j \varepsilon_0)^{1/2}$, $j = e$ or $h$. Accordingly, Drude permittivity function can be modified to include contributions from both free electrons and holes: $\varepsilon(\omega) = 1 - \sum_{j=e,h} \left( \omega_{p,j}^2 / (\omega^2 + i\gamma_j \omega) \right)$.

Because of the lower free carrier concentration in semiconductors, interband transitions usually occur at frequencies above those where free-carrier **intraband transitions** become important. This spectral separation of direct and SPP-mediated absorption bands helps to reduce dissipative losses in semiconductor plasmonics. Plasma frequencies for several commonly used metals and doped semiconductors previously reported in the literature [27,85–89] are shown in Figure 7a, and follow the scaling law $\omega_p \sim (n_e)^{1/2}$ (red solid line).

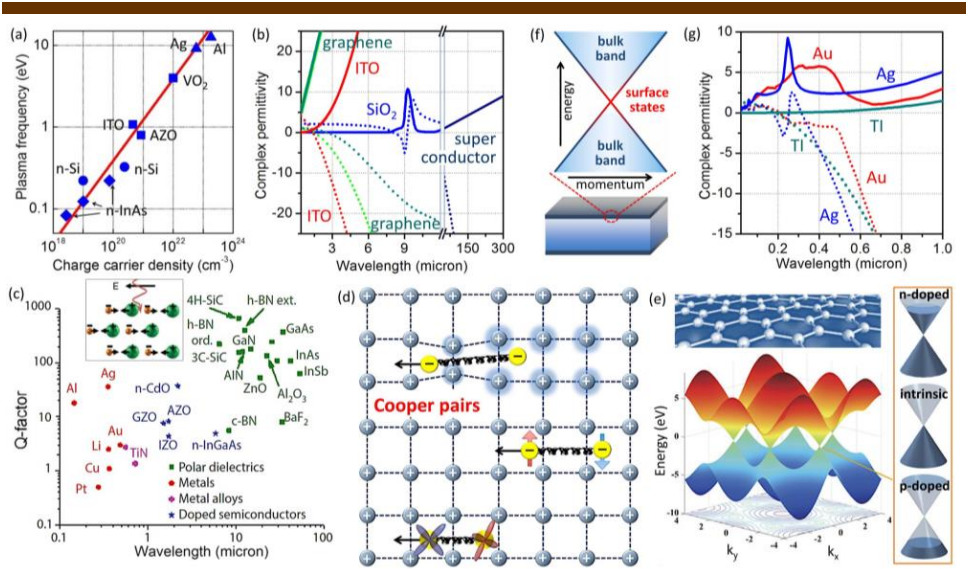

**Fig. 7. The search for low-loss plasmonic materials**. (a) Plasma frequencies of popular plasmonic materials scaling with the free charge carrier density. Reproduced with permission from [27], ©Elsevier. (b) Real (dotted lines) and imaginary (solid lines) parts of the permittivity of a doped semiconductor ITO (red) [89], polar dielectric $SiO_2$ [89] (blue), superconductor $YBa_2Cu_3O_7$ [90] (navy) and graphene at two different doping levels [91] (green & teal). (c) Quality factors of various plasmonic and polar materials in the ultraviolet-to-far-infrared frequency ranges. The top inset is a schematic illustrating the collective oscillations of atomic displacements in the form of optical phonons in polar dielectrics; modified from [92]. (d) The three types of mechanisms underlying the emergence of superconducting electrons: phonons (top), electron spin pairing (middle), and fluctuations between electron orbitals (bottom). (e) Band structure of a graphene monolayer. Insets: Dirac cones at different doping levels (intrinsic, *n*-doped, and *p*-doped), modified from [93], ©NPG. (g) Real (dotted lines) and imaginary (solid lines) parts of the permittivity of Au (red) and Ag (blue) noble metals as well as of the hypothetic Drude material that models the properties of the surface layer of topological insulator $Bi_{1.5}Sb_{0.5}Te_{1.8}Se_{1.2}$ (teal) [94]. (f) Schematic of the topological insulator bandstructure.

The complex permittivity of a popular plasmonic semiconductor – indium tin oxide (ITO) – is plotted in Fig. 7b (red lines) [89]. It can be clearly seen that unlike noble metals, this material exhibits plasmonic behavior in the infrared spectral range. Doped semiconductors have an additional advantage over noble metals because they can be spectrally tuned through the amount of doping, which shifts their plasma frequency. However, they still do not offer noticeable improvement in the reduction of dissipative losses over noble metals. This is illustrated in Fig. 7c, where the Q-factors of various metals (red dots) and doped semiconductors (blue stars) are plotted at the wavelength where $Re(\varepsilon) = -2$ for each material [92]. The Q-factor is defined as the ratio: $Q(\omega) = (\omega \cdot d\, Re(\varepsilon)/d\omega)/2\,Im(\varepsilon) \approx \omega/\gamma$. It can be estimated from the measured material permittivity function and is independent of the plasmonic nanostructure geometry. Larger Q-factors correspond to lower dissipative losses.

Figure 7c demonstrates that larger material Q-factors can be achieved in **polar dielectrics** such as silica, boron nitride, silicon carbide, etc [92,95]. Polar materials support **surface phonon polariton modes** (SPhPs), which stem from coupling of polar optical phonons and propagating photons (inset to Fig. 7c). These collective oscillations do not involve free charge carriers and thus suffer less from dissipative losses. However, applications of SPhP modes are limited to the mid-IR to **terahertz** (THz) spectral ranges, because they can only be excited within the so-called **Reststrahlen** frequency band between the longitudinal (LO) and transverse optic (TO) phonon frequencies, $\omega_{TO} < \omega < \omega_{LO}$. The TO and LO phonon modes correspond to out-of-phase atomic lattice vibrations with wavevectors aligned parallel (LO) and perpendicular (TO) to the incident field. In analogy to modeling the interband electron transitions in metals, the permittivity of the polar dielectric can be approximated via the Lorentz oscillator model: $\varepsilon(\omega) = \varepsilon_\infty \cdot \left(1 + (\omega_{LO}^2 - \omega_{TO}^2)/(\omega_{TO}^2 - \omega^2 - i\omega\gamma)\right)$. An example of the polar material complex permittivity function is shown in Fig. 7b (blue lines) for the case of silica ($SiO_2$) [89,96]. It can be seen that $SiO_2$ permittivity becomes negative within its Reststrahlen band, which is centered in the mid-IR spectral range. This significantly limits the spectral range where SPhP modes can be excited. Most other polar dielectrics exhibit polaritonic activity at even longer wavelengths (Fig. 7c, green squares) [74,80]. Overall, polar dielectrics offer an opportunity to achieve sub-diffraction confinement and optical losses significantly lower than those in noble metals and doped semiconductors, albeit only in the longer-wavelength spectral ranges.

Plasmonic response of metals, on the other hand, is not limited to the visible range. For example, silver exhibits plasmonic behavior at frequencies above a few THz and throughout the IR and visible spectral ranges [90]. As already illustrated in the previous section, this offers opportunities for tuning SPP resonances in metal nanostructures to the frequencies away from the direct absorption range. At low frequencies, charge carrier transport and thus optical response in metals is driven by the Ohm's law and not by collective dynamics of the electron plasma. In the low frequency range, **superconductors** have recently emerged as a promising THz plasmonic material platform that offers a new mechanism of dissipative losses reduction [90]. Superconductivity in a material arises when two electrons bind together into so-called **Cooper pairs** [97,98]. This pairing leads to a gap in the energy spectrum of the superconducting material, which makes the electrons insensitive to the mechanisms causing electrical resistance. Electrons can bind into Cooper pairs in different ways, leading to different categories of superconductors, which are illustrated in Fig. 7d. Pairing can be induced either by phonons (top Cooper pair in Fig. 7d) or by magnetic spin–spin interactions (middle pair in Fig. 7d). Recent work suggests that pairing can also be mediated by inter-orbital coupling (bottom pair in Fig. 7d) or a combination of orbital and spin fluctuations [99]. Collective plasma oscillations of superconducting electrons joined in Cooper pairs become immune to collision-induced damping, potentially paving the way to realization of low-loss THz plasmonic networks. The permittivity of the superconductor can be described via a two-fluid model, assuming existence of two non-interacting electronic sub-systems that contribute to the overall plasmonic response. These two systems include superconducting electrons (Cooper pairs), and non-paired electrons that are prone to scattering and the associated energy loss. A generalized Drude model can be used to model the local dielectric function of such a two-component electron plasma inside a superconductor: $\varepsilon(\omega) = 1 - (\omega_s^2/\omega^2) - \omega_p^2/(\omega^2 + i\gamma\omega)$ [83]. Here, the second term describes the contribution from the Cooper pairs with plasma frequency $\omega_s$, while the last term accounts for the conventional electron plasma with plasma frequency $\omega_p$ and damping parameter $\gamma$. The permittivity of $YBa_2Cu_3O_7$ superconductor approximated by using this two-fluid model is plotted in Fig. 7b (navy lines). It exhibits the imaginary part lower in magnitude than the real part, thus giving rise to SPP oscillations. Plasmonic response of superconductors is however limited to terahertz frequencies and below. At higher frequencies, superconductors become lossy as the energy of the incident photon is sufficient to break the Cooper pair and destroy superconductivity.

Another promising family of materials for the mid-infrared to terahertz frequency range are materials supporting collective plasmon-polariton oscillations of massless Dirac electrons [100]. **Dirac SPPs** can be observed in two-dimensional (2D) electron systems such as graphene monolayers shown in the upper inset to Fig. 7e. As a result of atomistic confinement of charge carriers, electron bandstructure of such materials exhibits symmetry-protected singularities known as Dirac points [101–104] (Fig. 7e). Linear energy dispersion near the **Dirac points** yields vanishing effective mass of charge carriers, high Fermi velocity, and a huge electrical mobility. A combination of these properties with the possibility of tuning the Fermi level by material doping (right inset to Fig. 7e) makes possible engineering tunable plasmon resonances for IR and THz applications [93,105,106]. The conductivity (and thus permittivity) of Dirac SPP materials such as graphene can be modeled as a combination of the interband and intraband contributions: $\sigma(\omega) = \sigma_{\text{intra}}(\omega) + \sigma_{\text{inter}}(\omega)$ [101,102]. However, similar to noble metals, at frequencies below the interband threshold, a simple Drude model that accounts for the intraband transitions serves as a reasonable approximation and agrees well with more advanced models [107]. The intraband contribution to the surface conductivity of graphene depends on frequency, dissipative damping $\gamma$ and the level of doping characterized by the chemical potential $\mu$:

$\sigma_{\text{intra}}(\omega) = i(2e^2 k_B T)/(\pi \hbar^2 \cdot (\omega + i\gamma)) \cdot \ln(2\cosh(\mu/2k_B T))$, where $\hbar$ and $k_B$ are the reduced Planck and the Boltzmann constants, respectively [91]. Figure 7b illustrates the changes in the graphene permittivity and spectral shifts of its plasma frequency at different levels of doping, i.e., for different $\mu$ (green and teal lines). Such tunability makes graphene a useful plasmonic material at IR and THz frequencies. Metals can also be used in the THz spectral range to form spoof plasmons on corrugated surfaces, with material Q-factors in theory approaching high values of $10^3$ [108].

The visible and ultraviolet (UV) parts of the optical spectrum, however, remain extremely challenging frequency bands for plasmonic applications. As previously discussed, noble metals such as gold and silver suffer from interband absorption losses in this range (Fig. 7f), while artificially doped semiconductors, polar dielectrics, superconductors, or graphene do not exhibit plasmonic response at these frequencies (Figs. 7b,c). A very interesting class of non-metals that can serve as alternative plasmonic materials in the visible range are semiconductors known as **topological insulators** (TIs). Topological insulators are materials that behave as insulators in bulk yet simultaneously exhibit high-mobility conducting electron states on their surfaces (Fig. 7g) [94,109–111]. The plasmonic response of TI materials can be attributed to a combination of surface optical conductivity of a nanoscale layer of topologically protected surface states and bulk optical conductivity due to the interband transitions in the material interior. While TI materials have been previously shown to exhibit THz plasmonic activity [112], recent work suggests that they can also provide a promising material platform for the visible spectral range [94]. The permittivity of the surface layer of a topological insulator $Bi_{1.5}Sb_{0.5}Te_{1.8}Se_{1.2}$ (BSTS) is plotted in Fig. 7f, and exhibits significantly lower losses than permittivities of Au and Ag. While dissipative losses within the underlying bulk of BSTS still negatively affect SPP propagation, frequency regions where BSTS offers lower losses for SPP propagation than noble metals have been experimentally identified [94].

It should be also noted that – unlike gold – many of the alternative low-loss plasmonic materials are compatible with the complementary metal oxide semiconductor (CMOS) technology. For example, metal-nitrides and graphene are CMOS compatible [86,113]. Transparent conductive oxides can be deposited at temperatures below 300°C, and thus can be integrated at final stages of the standard processing of silicon chips [114]. Topological insulators are also actively explored as alternatives for silicon for high-performance transistors [115].

While dramatic progress with identifying and synthesizing new plasmonic materials resulted in some interesting material discoveries [86], there is still much room for improvement of their dissipative losses. Plasmonic nanostructures and metamaterials still lag significantly behind their photonic counterparts based on isolators and semiconductors. This situation has recently led to the renewed interest in the development of all-dielectric nanostructures and metamaterials [71,116]. However, another alternative approach emerged that relies on the use of hybrid photonic-plasmonic networks [65,117–121]. Integration of SPP nanostructures with high-Q photonic elements into hybrid **optoplasmonic** platforms makes it possible to achieve extreme spectral and spatial localization of light simultaneously. This is illustrated in Fig. 8a, which compares quality factors of whispering-gallery (WG) modes of a high-Q $SiO_2$ microsphere with those of hybridized WG-SPP modes of the microsphere with an Au nanoparticle attached to its surface [120].

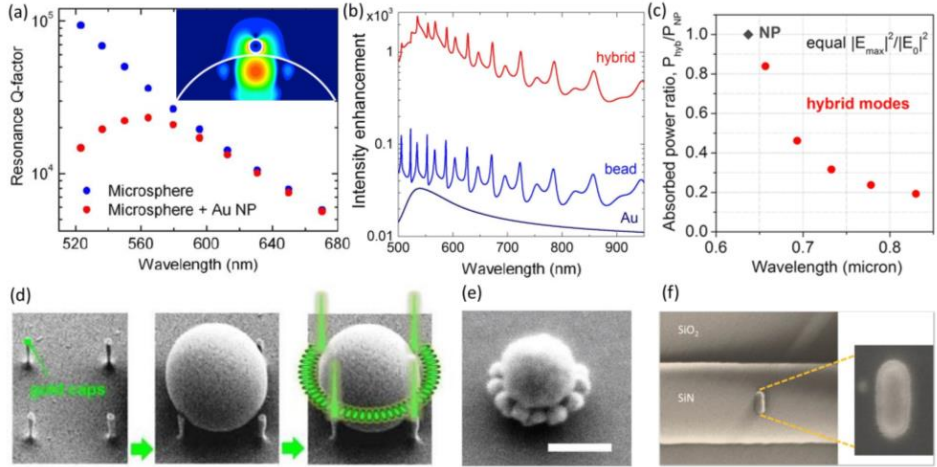

**Fig. 8. Dissipative loss reduction in hybrid optoplasmonic networks**. (a) Q-factors of WG modes in a $SiO_2$ microsphere (blue dots) and of hybridized WG-SPP modes in a $SiO_2$ microsphere – Au nanosphere dimer (red dots). The inset shows the electric field intensity distribution around the Au nanoparticle at one of the resonances. Reproduced with permission from [120], ©AIP. (b) Electric field intensity enhancement in the hot spot forming on the standalone Au nanoparticle (navy), a standalone $SiO_2$ microsphere (blue), and in a hybrid microsphere-nanoparticle dimer (red). (c) Optical power absorbed in the Au nanoparticle at the frequencies of hybridized WG-SPP modes in a microsphere-nanoparticle dimer normalized to the power absorbed in the standalone nanoparticle. The local field intensity at each resonance frequency is kept constant by adjusting the incident field intensity. Reproduced with permission from [122], ©ACS. (d-f) Hybrid optoplasmonic nanostructures fabricated by the template-assisted self-assembly (d,e) and e-beam lithography (f). Reproduced with permission from [119], ©ACS, and [123] ©OSA, respectively.

The Q-factors of the hybridized modes with frequencies outside of the range of direct transitions in Au are almost identical to those of WG modes of a standalone microsphere. Furthermore, Q-factors of the hybridized modes in the range of Au direct absorption remain orders of magnitude higher than the Q-factor of an SPP dipole mode of standalone nanoparticle [120,124] (see also Fig. 8b). The inset to Fig. 8a illustrates strong spatial light localization around the metal nanoparticle.

Higher mode Q-factors translate into larger local field intensity enhancements in hybrid optoplasmonic structures (Fig. 8b). Intensity enhancement in a hybrid dimer at frequencies corresponding to the excitation of hybridized WG-SPP modes significantly exceeds those of either the standalone nanoparticle or the standalone microsphere. In fact, this enhancement can significantly exceed the multiplicative effect of the SPP and the WG-mode enhancement owing to the formation of local powerflow features such as optical vortices [49,125]. The high local field intensity in optoplasmonic structures makes them promising high-figure-of-merit platforms for optical sensing and spectroscopy. In the frame of the first-order perturbation approximation, wavelength shift of a mode caused by a small molecule with a real polarizability $\alpha$ at position $\mathbf{r}_0$ is directly proportional to the field intensity at the molecule position $|\mathbf{E}(\mathbf{r}_0)|^2$ and inversely proportional to the energy density integrated over the whole hybridized mode volume $V$:

$\Delta\lambda/\lambda \sim (\alpha/\varepsilon_0 |\mathbf{E}(\mathbf{r}_0)|^2) / 2\int_V \varepsilon(\mathbf{r})|\mathbf{E}(\mathbf{r})|^2 dV$ [120,121]. Significant increase of the field intensity in the hybrid dimer (Fig. 8b) translates into its larger detection sensitivity. Furthermore, orders-of-magnitude enhanced Q-factors of hybridized WG-SPP modes also increase the hybrid sensor spectral resolution and thus the overall figure of merit.

The enhanced field intensity in optoplasmonic platforms is naturally accompanied by the increase of their absorption cross-sections over those of individual photonic and plasmonic components. However, to achieve the same local field intensity on the nanoparticle integrated into a hybrid optoplasmonic structure as on the standalone nanoparticle, optical pump irradiance can be reduced by orders of magnitude. This saves power and significantly lowers the power absorbed in other areas of the optical chip. The absorbed power can be further reduced by tuning the hybridized modes resonances away from the direct absorption frequency range. This is illustrated in Fig. 8c, which shows the relative power absorbed in an Au nanoparticle coupled to a $SiO_2$ microsphere at the frequencies of hybridized WG-SPP modes progressively shifted away from the direct absorption range [122].

Hybrid optoplasmonic networks can be formed by using a variety of fabrication techniques, including a combination of electron beam lithography and self-assembly (Fig. 8d) [118,124,126,127], template-assisted self-assembly (Fig. 8e) [119], multi-step lithography (Fig. 8f) [123], and a combination of self-assembly, optical fiber melting and piezoelectric coupling control [120,121,128]. Photonic components that can be incorporated into hybrid optoplasmonic networks range from micro- and nanospheres, microdisks, microrings, Fabry-Perot cavities and integrated optical waveguides to photonic crystal cavities and optical fibers [65,129–136]. Finally, the increased local density of optical states in hybrid optoplasmonic structures provides significant enhancement of the radiative rate of quantum emitters embedded into hybrid networks [117,126,137,138]. This ultimately opens up opportunities for using hybrid optoplasmonic structures with deep subwavelength optical confinement to achieve optical amplification and lasing action [139,140].

## 5. MITIGATING THE EFFECT OF PLASMONIC DISSIPATIVE LOSSES

As discussed in section 2, dephasing of SPP resonances through intrinsic damping processes, including SPP waves scattering on phonons, lattice defects, and interfaces as well as Landau damping, creates excited electron-hole pairs. These energetic charge carriers lose energy and thermalize via electron−electron scattering and electron−phonon coupling, with the excess energy ultimately dissipating to the environment via the processes of thermal radiation and heat conduction (Fig. 9). A photo-excited carrier relaxes initially through two competing pathways: carrier-carrier scattering and optical phonon emission. In the former process, the energy of **photoexcited 'hot' carriers** remains in the electron system, and is gradually transferred to 'cold' carriers that gain extra kinetic energy. At the same time, the energy lost due to electron-phonon scattering is transferred to the material crystal lattice in the form of heat, thus raising local temperature. Although these processes occur on the ultra-fast time scale (Fig. 9), the energy of the hot charge carriers can still be partially harvested before it completely dissipates as heat [19,141–144]. The approaches to harvest the SPP energy include plasmon-driven resonant energy transfer [145] and hot carriers collection via the processes of **internal photoemission** or **tunneling** through the metal-semiconductor barrier [27,141,144,146–150].

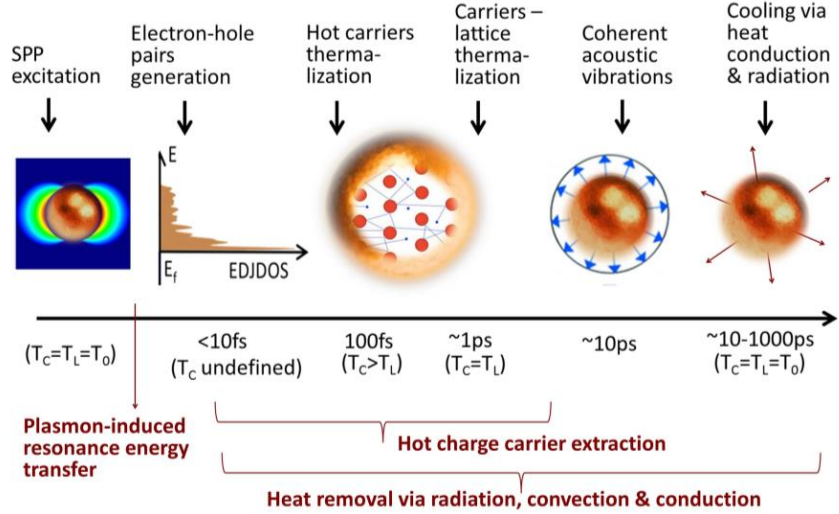

**Fig. 9. Ultra-fast evolution of the SPP energy dissipation through various channels.** Various mechanisms of SPP energy extraction and local heating mitigation are shown at the bottom.

The process of hot carriers harvesting via internal photoemission through the **Schottky barrier** between an Au absorber and a semiconductor is illustrated in Fig. 10a. The Schottky junction that forms at the interface between metal and n-type semiconductor can act as a frequency-selective filter to extract hot electrons generated in the process of either direct photon absorption or the SPP decay. Alternatively, the junction between metal and p-type semiconductor can be used to harvest hot holes. A simple estimation of the limiting responsivity of the Schottky junction for the case of direct photon absorption can be performed by using the concept of the joint density of states, $D(E,\hbar\omega) = \rho(E-\hbar\omega) \cdot f(E-\hbar\omega) \cdot \rho(E) \cdot (1-f(E))$ [146,151]. $D(E,\hbar\omega)$ quantifies the number of initial $\rho(E-\hbar\omega)$ and final $\rho(E)$ electronic states in the material that are available for direct transitions given the absorbed photon energy. Here, $f(E)$ is the Fermi-Dirac distribution function, which defines the occupancy probability of an available energy level at a given temperature $T$. Figure 10a shows the density of available electron states (blue) and $D(E,\hbar\omega)$ for both electrons (red) and holes (orange) in Au, calculated using the ab-initio Density Functional Theory (DFT) [89,146]. The responsivity limits of Schottky devices can be calculated assuming that all the hot carriers with energies higher than the Schottky barrier height $\Phi_B$ are collected (Fig. 10b) [146].

Dominant contribution from the excited *d*-band electrons to the hot electron population in Au is extremely unfavorable for achieving high responsivity (Figs. 10a,b). It should be noted that the responsivity of realistic devices is significantly lower due to imperfect absorption and hot carriers thermalization losses, which reportedly are especially severe for hot holes arising from the *d* states [144].

Interestingly, the Landau damping mechanism, which is typically detrimental for achieving low-loss SPP propagation, can actually play a positive role in applications of plasmonics related to hot carrier generation and harvesting. It has already been demonstrated both theoretically and experimentally that hot carriers generated in noble metals via plasmon decay can be much more energetic than those resulting from direct photoexcitation [19,144,148]. This positive effect stems from the differences in the physical mechanisms underlying electron–hole pair generation via SPP decay and direct absorption (Fig. 6b). SPP decay induces indirect intra-band electron transitions and allows excitation of electrons from the energy levels just below the Fermi energy. This results in generation of substantially higher-energy hot electrons [144]. These electrons can pass through the Schottky barrier, yielding much higher device responsivity under the same illumination conditions.

The effect of the SPP-induced decay in the generation of hot electrons is illustrated in Fig. 10c, which compares measured responsivity of a simple Schottky device shown in the inset under either direct absorption or plasmon excitation scenarios. The two scenarios are segregated by adjusting the polarization state of the incident light. The illumination by light with the transverse electric (TE) polarization (i.e. with the electric field oscillating perpendicular to the metal gratings) results in the SPP excitation. In turn, the SPP decay lead to the significant increase of the measured device responsivity (compare solid and dashed lines in Fig. 10c) [148].

Non-equilibrium charge carriers excited by light in plasmonic nanostructures also offer an opportunity to achieve upconversion of photon energy. Hot electrons and holes excited by low-energy photons and injected into an adjacent semiconductor quantum well can radiatively recombine to emit a photon of higher energy [152]. Emerging applications of plasmonics for hot electron harvesting, photocatalysis, and ultrafast switching [27,106,143,153–156] underscore the high potential of turning the plasmonic losses into performance gain.

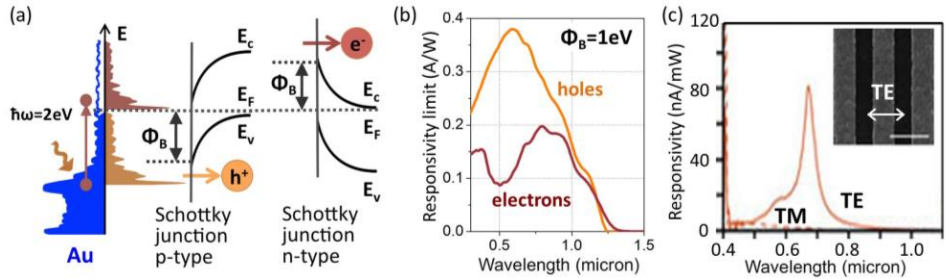

**Fig. 10. Collection of hot charge carriers via internal photoemission through the Schottky barrier between a metal and a semiconductor.** (a) Population of available electron energy states in the dark at T=300K (blue area) and after the direct photo-excitation by photons with the energy $\hbar\omega = 2eV$ (red area) in Au [146]. The Au electron DOS is calculated by using the DFT method. The hot holes population is shown in orange. (b) Responsivity limits of Schottky junctions as a function of the absorbed photon energy for hot electron (red) and hot holes (orange) harvesting schemes. (c) Measured responsivity of the Schottky device in the case of the SPP-induced absorption under TE-polarized illumination (solid line) and direct absorption under TM-polarized illumination (dashed). The Schottky device SEM image is shown in the inset. Adapted with permission from [148], NPG.

Thermalization of the SPP-excited charge carriers with the lattice results in the localized heating of plasmonic elements and their immediate surroundings (Fig. 9) [21,27,157,158], which may further increase SPP absorption losses [159] and/or lead to the thermally-induced material failures [160]. Plasmonic nanostructures are often smaller than both the phonon mean-free path in their host or substrate materials and the wavelength of thermal emission. They are also comparable in size to the mean free path of the air molecules. Accordingly, plasmonic nanostructures do not cool down in the same way bulk materials do due to a combination of extremely low thermal emittance in the mid-to-far infrared spectral range [122] and reduced conductive and convective heat transfer to the environment [161]. The local heating can affect performance of plasmonic sensors, tweezers, and plasmon-enhanced photovoltaic (PV) cells [162–165]. A temperature increase as low as 1 K reduces efficiency of crystalline Si solar cells by approximately 0.5%, while heating of plasmonic nanotweezers [166,167] limits the maximum optical power thus restricting the magnitude of the optical force that can be achieved [168]. Photo-induced heating of plasmonic nanoantennas can also cause their melting and morphology changes [169]. Melting reduces the sharpness of the nanoscale features and ultimately turns plasmonic particles of any geometry into nanospheres. Such morphology changes modify the spectral response of SPP elements and reduce the achievable near-field enhancement [169]. Furthermore, for applications in sensing, spectroscopy, and near-field imaging, excessive localized heating might be harmful to the material or tissue that is being probed [170].

To alleviate the effect of plasmonic heating, the excess thermal energy needs to be removed from the plasmonic chip via conductive, convective or radiative channels. Active cooling technologies demonstrated for plasmonic heat management in optical chips and PV cells include active air ducts cooling [171], water cooling [172], and heat-pipe cooling [173]. An example of the latter cooling approach is illustrated in Fig. 11a, which shows how a heat-pipe plate integrated with a plasmonic PV cell can be used to efficiently transfer heat away from the hot spots by making use of the liquid-vapor phase change of the working fluid. This active cooling approach has led to the experimental demonstration of the operating temperature reduction of the plasmonic PV cells by over 16°C (Fig. 11b), with major implications for the PV cell efficiency [173].

However, less technologically complex passive cooling solutions might be preferable for the plasmonic hot spots mitigation in optical and electronic systems. A simple solution to alleviating heating in plasmonic sensors is to use either nanohole arrays in metal membranes or arrays of nanopillars on metal substrates integrated with a heat sink underneath [168]. This facilitates heat removal via conduction through the metal volume and helps to alleviate local heating effects on individual plasmonic nanoantennas. Fig. 11c shows an example of a passive cooling system designed for high-performance plasmonic chips that makes use of enhanced heat conduction and natural convection [174]. Modeling predicts that multi-layered thermal interfaces of nano- and micrometer thickness, combined with simple convective cooling systems, can help to reduce the temperature of the plasmonic chip by several hundred degrees [173].

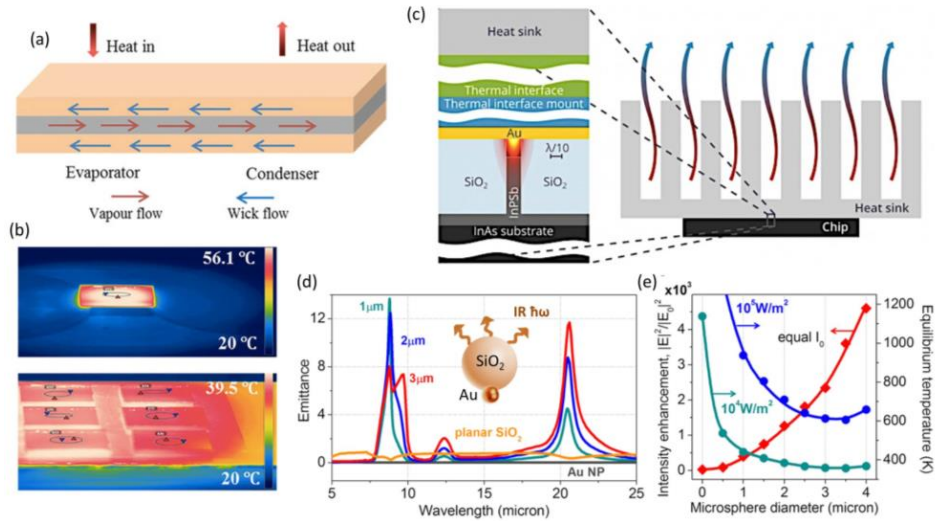

**Fig. 11. Mitigation of the SPP-induced heating via heat conduction, convection and radiation.** (a) Working mechanism of the heat-pipe plate for active heat removal via conduction and fluid convection. (b) Thermal images illustrating stabilized temperature distributions for the plasmonic PV cells with (bottom image) and without (top image) the integrated heat-pipe plate. Images in (a,b) are reproduced with permission from [173], ©NPG. (c) Schematic of an electrically pumped active T-shaped plasmonic waveguide integrated with the heat sink for passive convective cooling via a series of thermal interface layers. Reproduced with permission from [174], ©ACS. (d) IR effective emittance of hybrid optoplasmonic dimers composed of an Au nanosphere and a $SiO_2$ microsphere of varying diameter. The emittance spectra of a $SiO_2$ surface and of a standalone Au sphere are shown for comparison as orange and gray lines, respectively. (e) Equilibrium temperature and the near-field intensity enhancement of optoplasmonic dimers as a function of the $SiO_2$ microsphere size. Diameter value 0 corresponds to the case of the standalone Au nanosphere. Panels d,e are reproduced with permission from [22], ©ACS.

Finally, passive radiative cooling can be a powerful mechanism for temperature reduction [175,176], especially in the situations where other heat removal channels are suppressed. However, one of the reasons why metal plasmonic nanoantennas overheat under light illumination is their extremely low thermal emittance in the mid-to-far-infrared frequency range [122]. Nevertheless, simulations predict that optoplasmonic nanostructures made of metals and polar dielectrics (see Fig. 8) can simultaneously offer orders-of-magnitude intensity enhancement and local temperature control [122]. It was already discussed (Fig. 8c) that enhanced light focusing in hybrid optoplasmonic structures yields high local field enhancement at significantly lower powers of the optical pump than in their purely plasmonic counterparts. This reduces the power absorbed in plasmonic nanoparticles as well as in the other parts of the optical chip. Optoplasmonic structures can also be designed to have strong resonant interactions with photons not only in the visible but also in the mid-to-far infrared wavelength ranges via excitation of surface phonon-polariton modes in polar dielectrics. Excitation of SPhP modes in dielectric particles yields infrared emittance cross-sections exceeding their geometric cross-sections (Fig. 11d) [29]. This high infrared emittance leads to the enhanced passive cooling through thermal radiation (Fig. 11e) [22]. Modeling also predicts that enhanced convective cooling of hybrid optoplasmonic structures can further contribute to their temperature reduction [22].

## 6. USING LOCALLY GENERATED HEAT

Recently, many new applications of plasmonics have emerged that, instead of attempting to alleviate plasmonic losses,

> Using locally generated heat opens new application horizons for plasmonics, including solar and thermal energy harvesting, nanofabrication, nano-manipulation, heat-assisted magnetic recording and cancer therapy.

make use of heat locally generated on plasmonic nanostructures. These applications exploit the excellent photo-thermal conversion efficiency of plasmonic materials to achieve targeted delivery of both light and heat to sub-diffraction-limited nanoscale areas, opening the door to the new field of **thermoplasmonics** [177].

### 6.1 Nanofabrication and nano-recording

Photo-induced heating of plasmonic nanoparticles can change their morphology due to the melting of metal [73,169]. Surface melting of nanoparticles and nanostructures can occur at temperatures much lower than the melting temperature of the same material in the bulk form. Melting reduces the sharpness of the nanoscale features and ultimately turns particles of any shape into nanospheres once high enough temperature is reached. Accordingly, post-fabrication heat reflow process can be used to correct for fabrication imperfections and to ensure that all the plasmonic nanoparticles are uniform and have smooth surfaces. An example of the heat reflow post-fabrication treatment is shown in Fig. 12a, where the gold nanodisks fabricated by the electron beam lithography process have been converted into uniform and smooth spherical particles [73].

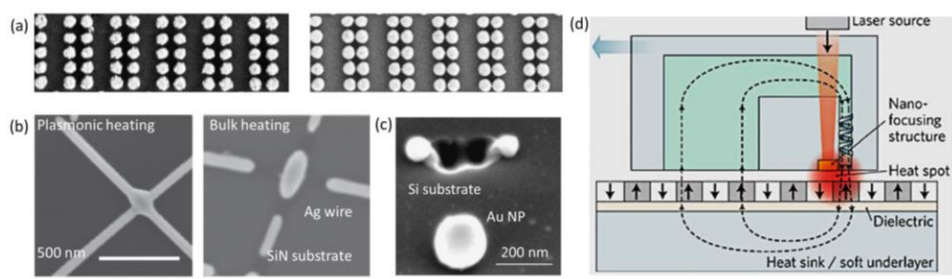

**Fig. 12. SPP-induced heating for nanofabrication and heat-assisted magnetic recording.** (a) Heat reflow treatment converts rough Au nanodisks into smooth Au nanospheres. Reproduced with permission from [73], ©ACS. (b) SPP-driven self-limiting nanowelding delivers heat locally and does not damage the substrate. Reproduced with permission from [178], ©NPG. (c) Nanoholes drilled in silicon

via plasmonic heating of Au nanoparticles. Modified with permission from [179], ©Elsevier. (d) Schematic of the heat-induced magnetic recording technique that makes use of localized SPP heating to reduce the coercivity of the magnetic medium. Reproduced with permission from [180], ©NPG.

For plasmonic nanostructures fabricated on heat-resistant substrates, thermal reflow can be performed by uniformly heating the whole plasmonic chip. However, local heat application is often required to avoid damaging the substrates with low melting temperatures such as polymers, and to perform high-precision procedures as nanowelding [178] (Fig. 12b) or nanoholes drilling [160,179] (Fig. 12c). Figure 12b illustrates an SPP-induced nano-welding technique, which be used to connect metallic nanowires into complex on-chip networks [178]. Highly-localized SPP modes excited by laser light at nanowire junctions enable effective light concentration and nano-localized heating, which helps to join the nanowires more efficiently than bulk heating and does not damage the underlying substrate. Furthermore, as the SPP local field intensity is highly sensitive to the nano-junction geometry, the welding process can be self-limited and stops once the inter-wire connection forms. A similar plasmonic nano-welding process has been successfully used to locally weld polymer nanofibers coated with Ag nanoparticles [181].

Furthermore, local heating of nanoscale metallic catalysts can be used to initiate and control growth of semiconductor nanowires and carbon nanotubes [182]. It has been shown that local plasmonic heating can be orders of magnitude more energy efficient than the conventional approach of heating the entire substrate. The substrates can also be locally modified via localized heating induced by plasmonic nanoparticles. An example of the nanohole drilled into a silicon surface by femtosecond laser pulse irradiation mediated by an Au nanoparticle is shown in Fig. 12c [160,179]. Here, plasmonic heat localization enabled fabrication of sub-diffraction-size nanoholes with diameters of about 150 nm by laser irradiation with the wavelength 820nm at fluences below the threshold for Si surface thermal modification [160,179].

Nanoscale plasmonic-induced heating has also revolutionized the data storage technology. To increase storage density of hard disk drives, the magnetic grain size needs to be reduced to the nanoscale footprint. Although materials with large magnetic anisotropy can support grains as small as 2 nm in size, the coercivity of these materials is greater than the magnetic field that can be generated by a recording head. Heating the magnetic medium above its Curie point during recording can help to reduce its coercivity to zero. Plasmonic heating makes this process possible as it enables delivery of heat to a spot much smaller than the diffraction limit of light to avoid heating neighboring tracks during the writing process [183–185]. The **heat-assisted magnetic recording** (HAMR) technology [180] makes use of localized plasmonic heating (Fig. 12d) and holds promise to increase the data storage density above 10 Tb/in$^2$.

Finally, it should be noted that both radiative and dissipative plasmonic loss mechanisms (see Fig. 2) can be used in the process of color and art creation. Strong scattering at select frequencies corresponding to the excitation of the dipole SPP modes on plasmonic nanoparticles results in the structural color formation that can be visible by a naked eye. This effect has been used for centuries to make colored glass by embedding metal nanoparticles into windows and glassware. One of the most famous examples of plasmonic structural color in art is shown in Fig. 13a. A Roman Lycurgus Cup – a 4th-century Roman glass cage cup – exhibits different colors depending on the light illumination direction. The underlying mechanisms of such color dichroism are spectrally-selective dissipative and radiative losses induced by light interaction with Au and Ag plasmonic nanoparticles embedded in glass. As the SPP resonances are extremely sensitive to the particle shapes and sizes, the color can be controlled by tuning these parameters.

It has been recently realized that localized thermal reflow driven by plasmonic dissipative losses can be used for targeted modification of nanoparticle shapes (Fig. 13b and also Fig. 12a). Shape modifications in turn result in the change of the structural color formed by light scattering on nanoparticles. Figure 13b reveals how gradual rounding of Au triangular nanoparticles as a result of plasmonic heating under laser illumination yields visible color change [169]. Depending on the laser pulse energy density, different surface morphologies that support different plasmonic resonances leading to different color appearances can be created pixel by pixel [186] (Fig. 13c).

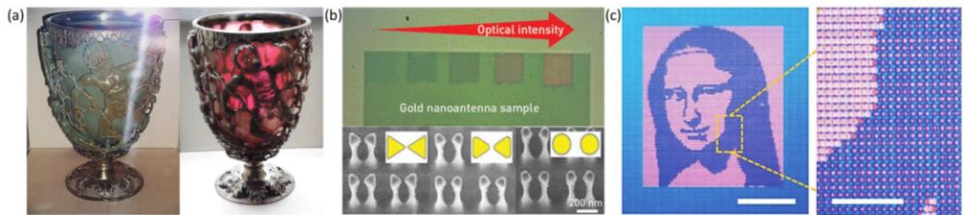

**Fig. 13. From macro- to nano-scale plasmonic color generation via radiative and dissipative plasmonic loss mechanisms.** (a) The 4th-century Roman Lycurgus Cup colored by light scattering on glass-embedded Au and Ag nanoparticles (image credit: Johnbod, Wikimedia Commons). (b) Nanoscale optical writing via localized plasmonic heating and material melting yields visible color modification of the plasmonic chip. Modified with permission from [169], ©ACS. (c) Pixel-by-pixel SPP-induced thermal reflow of Au nanoparticles yields plasmonic paintings with nanoscale features. Reproduced with permission from [186], ©NPG.

### 6.2 Photothermal therapies, spectroscopy and nano-manipulation

The unique localized nature of plasmonic heating of metal nanoparticles has also been harnessed for thermal destruction of cancerous cells both *in vitro* and *in vivo* [187–193]. Au nanoparticles in particular have been extensively studied for this application owing to their high photothermal heating efficiency and ease of surface functionalization [188]. The absorption cross-sections of Au nanoparticles are several orders of magnitude larger than the strongest absorbing organic chromophores. Furthermore, the SPP resonances can be easily tuned to a specific laser wavelength in the so-called "water window" (700–1200 nm), in which body tissues are more optically transparent to laser light, making possible *in vivo* treatment of tumors. The significant difference between traditional hyperthermia and **plasmonic photothermal therapy** is that plasmonic heating only affects the area in the close vicinity the gold nanoparticles (Fig. 14a). While the local temperatures can rise up to tens or hundreds of degrees above physiological temperature on very short timescales, the surrounding healthy tissue is not affected, potentially reducing negative side effects of therapy.

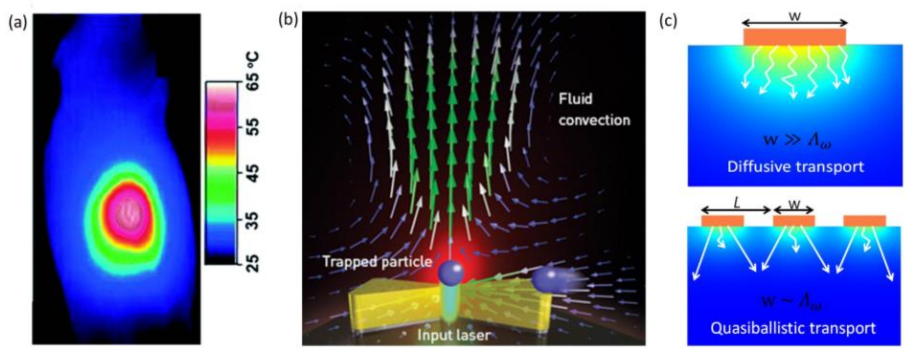

**Fig. 14. Plasmonic heating for cancer photothermal therapy, nano-manipulation and materials spectroscopy.** (a) Thermographic images of tumor-bearing mice showing heating localization during the photothermal therapy. Adapted with permission from [194], ©Wiley. (b) Schematic of plasmonic nanotweezers, with convection currents drawing particles into the trapping volume on the plasmonic bow-tie antenna. Reproduced with permission from [195] ©OSA. (c) Localized heating for the phonon

spectroscopy of materials. Heat transport in semiconductors is diffusive when the heater size *w* is much larger than phonon MFPs $\Lambda_\omega$ (top) and quasiballistic when heating is localized to lengthscales comparable with phonon MFPs (bottom). Adapted with permission from [196], ©NPG.

Interestingly, localized plasmonic heating used in tumor ablation can also provide a remote temperature readout mechanism through the process of photon upconversion [197,198]. Temperature changes affect the efficiency of the anti-Stokes photon emission of luminescent materials such as organic dyes, polymers, and nanophosphors. Measured remotely, photon emission can provide a useful *in vivo* readout of local temperature on the nanoparticle. Temperature-feedback luminescent materials combined with plasmonic nanoparticles already find use in real-time monitoring of microscopic temperature to avoid causing damage to surrounding benign cells during photothermal therapy [170,189,199,194]. Finally, SPP excitation on nanoparticles offers a possibility to combine nanoscale imaging and targeted therapy delivery [200–202].

Another avenue of research that benefits from combining the strong near-field light intensity and strong local temperature gradients is molecular sensing and nano-manipulation. The subwavelength confinement and strong SPP-driven enhancement of local electric field enables the use of plasmonic nanoantennas as **nanotweezers** for optical trapping. In addition to providing local field enhancement, heat generation by optically-excited plasmonic nanostructures can drive local fluid convection (Fig. 14b). Local convection helps to draw target nanoparticles or molecules into the trapping volume surrounding the illuminated spot [203–205,195]. Local convective currents induced by plasmonic heating can find uses in many other optofluidic applications for particles sorting, fluid mixing, etc.

Ultrafast optical spectroscopy coupled with plasmonic heating approach has also led to significant progress in measuring heat carrier's mean free path distributions in many material systems. In most semiconductors and dielectrics, thermal transport is a broadband process that involves cumulative contributions from phonons spanning a wide range of MFPs [206,207]. To experimentally extract the phonon MFP distribution, optical spectroscopy can be used to observe **quasiballistic thermal transport** created through localized heating of materials on progressively smaller lengthscales [208]. As shown in Fig. 14c (top), phonons experience sufficient scattering to establish local thermodynamic equilibrium when the heater size is much larger than the phonon MFPs. This results in diffusive transport that can be accurately described by the classical heat diffusion theory. In contrast, quasiballistic transport is induced when the heater dimension is comparable with or smaller than the phonon MFPs, as shown in Fig. 14c (bottom), since long-MFP phonons do not undergo scattering on small nanostructures. Systematically varying the characteristic heater lengthscales across a wide range gives information on the relative contribution of different phonon MFPs to thermal transport. The hot-spot size that can be generated via photon absorption in the material is ultimately diffraction-limited by the optical wavelength. However, light and heat localization on plasmonic nanostructures helps to meet the non-trivial challenge of accessing the quasiballistic phonon transport regime to map the MFP distributions of various materials [209,210,196].

Plasmonic heating can also be used for **dynamic thermal modulation** of material optical properties. Many materials exhibit large changes in their electrical and/or optical properties with temperature, including phase transitions, refractive index changes and carrier density variations [211]. For example, optically activated nanoscale plasmonic heat sources have been used to induce gel-fluid phase transitions in phospholipid membranes [212]. Polymer brushes have been actuated and probed through localized plasmonic heating that induces reversible phase transitions, thus allowing controlled transition between a collapsed conformational state (i.e. polymer shrinking) and an extended conformational state upon cooling (i.e. polymer stretching) [213,214]. Gold nanoparticles have also been used as fine tools to control heat-induced polymerization reactions at the nanoscale, with applications for the controlled synthesis of polymer nanoparticles and nanowires [215].

Thermally tunable phase change materials, including $VO_2$ [216–223], superconductors, and $SrTiO_3$ [224] can be incorporated into metamaterials to achieve dynamical thermal modulation of their optical properties. Dynamic modulation can result in near-perfect-absorption to near-perfect-reflection switching [221], plasmon-induced transparency activation [219], dynamic antenna operation [218,225] and optical filtering [211]. While heating of phase change materials via substrate of the device is the conventional method of temperature tuning, plasmonic heating enables phase transitions to occur on a picosecond time scale enabling fast thermal switching of metamaterial response [220,225]. Semiconductors allow for modulation of their optical properties via a thermo-optic effect that leads to variations in bandgap edge absorption and material refractive index [226–228]. Thermal modulation of semiconductor properties is rarely used as it is considered too slow for modern telecommunications applications [229]. However, plasmonic heating of silicon optical resonators can yield modulations on time scale of hundreds of nano-seconds [227], which is faster than typical thermal-modulated optical devices. Modeling also predicts modulation speeds on the order of GHz to be achievable with small plasmonic nano-heaters in silicon [230], much faster than state-of-art kHz or MHz thermo-optic modulation.

### 6.3 Energy generation, storage and sustainability

Many applications in the renewable energy generation and storage can also benefit from the plasmonic heating mechanism. In particular, frequency-selective photon absorption and the excellent photo-thermal conversion efficiency of plasmonic nanoparticles makes them promising candidates for nanofluid filters for hybrid solar photovoltaic/thermal (PV/T) applications [155,231]. Hybrid PV/T systems combine a hot thermal absorber and a PV cell, and are designed to harvest the whole solar spectrum. Nanoparticles optimized to absorb only photons below the PV cell bandgap and photons with energies much higher than the bandgap can act as a selective optical filter, passing a narrow range of photons to the PV while absorbing the rest (Fig. 15a). They simultaneously function as heat exchangers, efficiently conducting the absorbed heat to the working fluid.

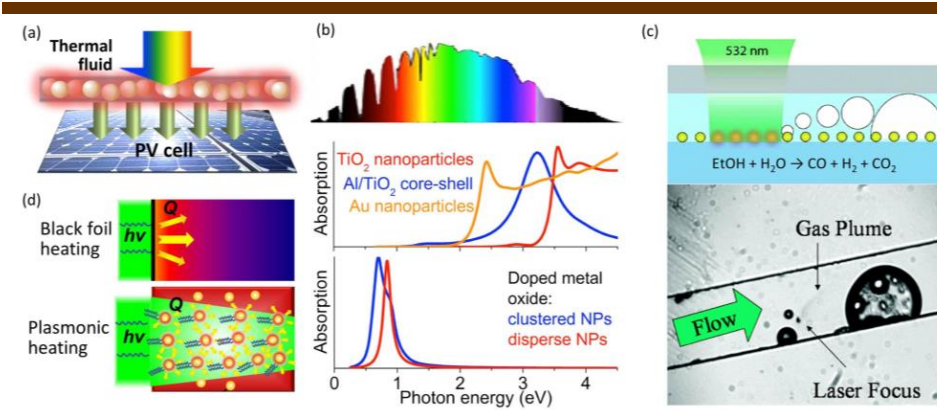

**Fig. 15. Plasmonic absorption and heating for applications in energy generation and storage.** (a) A schematic of a hybrid solar photovoltaic/thermal energy conversion platform with a frequency-selective plasmonic nanofluid absorber. (b) Noble metal plasmonic nanoparticles can harvest the high-frequency photons of the solar spectrum, while metal oxide materials such as Indium Tin oxide (ITO) or Aluminum-doped Zinc Oxide (AZO) can harvest the low-frequency photons. (c) Plasmon-driven thermocatalytic reaction. Reproduced with permission from [232], ©ACS. (d) Schematic of charging thermal energy storage materials through (top) a slow, conventional process of diffusive heating from a substrate and (bottom) instant plasmonic heating via SPP activation on dispersed Au nanoparticles. Reproduced with permission from [233] ©NPG.

A typical PV/T system schematically shown in Fig. 15a provides a high temperature fluid for storage or power production via a heat engine, and electricity generated in the PV cell, which operates at near-ambient temperature. Hybrid PV/T platforms can be theoretically designed to operate at near 80% in combined efficiency [234], which requires efficient spectral separation of photons. Tunable optical properties of nanofluid filters can be achieved by dispersing properly designed plasmonic nanoparticles in conventional thermal fluids. Figure 15b shows possible combinations of nanoparticles that can absorb the high-energy and the low-energy parts of the solar spectrum, while being very transparent to photons efficiently harvested by a silicon PV cell. The nanofluid filter bandwidth can be tuned by using other core-shell nanoparticles [235].

Localized plasmonic heating can also help to miniaturize many types of heterogeneous catalysis processes, which are used in fabrication of 90% of chemical products consumed by our society. While plasmon-enhanced photocatalysis has already attracted a lot of research efforts [236–238], **plasmon-enhanced thermocatalysis** – where localized heat is used to drive the chemical reaction – is more recently becoming a focus of intense study. Heterogeneous catalysis typically makes use of metal nanoparticles purely for their catalytic activity, however, these nanoparticles can be simultaneously used as photo-thermal transducers to provide the necessary heat of reaction under light illumination. One example of plasmon-driven thermocatalytic reaction is shown in Fig. 15c, where a liquid mixture of ethanol and water flows over gold nanoparticle catalysts in a microfluidic channel to produce $CO_2$, CO, and $H_2$ [232]. Localized plasmonic heating provides the heat of reaction to vapors present on the nanoparticle surfaces, yet allows the chip and the fluid to remain at room temperature. This local heating approach can be used with a variety of endothermic catalytic processes involving nanoparticles and can be performed on heat sensitive substrates. It has other potential benefits over conventional bulk thermocatalysis, including reduction of system thermal mass, precise spatial and temporal control of reaction [239] as well as rapid reaction quenching [240]. Furthermore, the combination of light and heat localization makes plasmonic heating approaches viable for solar-driven reduction of $CO_2$ and other practical solar-thermal catalytic applications. For example, plasmonic excitation of Au nanoparticles using low power illumination has already been used to achieve conversion of $CO_2$ and $H_2$ reactants to $CH_4$ and CO products [241].

Plasmon-mediated local heating can also modify the chemical vapor deposition (CVD) processes to selectively deposit materials of many different types without heating the substrate [242]. A huge advantage of the plasmonic optical heating is that it is localized not only in space but also in time, allowing controlled deposition onto different non-diffraction limited spots on the substrate. The local heating strategy is often orders of magnitude more energy efficient than conventional CVD processes that require heating an entire substrate. By locally heating nanoscale metallic catalysts, precise growth of semiconductor nanowires and carbon nanotubes has been successfully demonstrated down to the single nanostructure level in room-temperature chambers [182].

As already discussed, controlled deposition of heat can also be applied to induce phase transition in materials. Phase transitions from solid to liquid can be used to store thermal energy, making them important for energy and sustainability applications. Rather than inducing phase change through slow thermal diffusion from a hot surface to the bulk (Fig. 15d, top), nanoparticles dispersed in the bulk can be used for fast and uniform optical charging of the phase-change material (Fig. 15d, bottom [233]). Other phase transitions driven by plasmonic heating include solar evaporation and boiling of nanofluids [243–247], melting of ice using embedded nanoparticles [248], and formation of micro- and nano-bubbles [249–252] (see e.g. Fig. 15c).

In particular, plasmonic heating is useful in solar vapor generation for applications such as water desalination and sterilization. High temperature steam was generated recently by using a gold-silica core-shell nanofluid to absorb concentrated solar flux [253]. In follow-up efforts, a solar-powered autoclave useful for sterilizing medical instruments in developing countries was developed [245]. Additional performance increases are possible by tuning the plasmonic optical properties for heat and light localization on the interface between fluid and air. This can be achieved through engineering both radiative and absorptive plasmonic loss mechanisms [158,254]. For example, it was found that light scattered by nanoparticles is absorbed closer to the surface of the liquid (Fig. 16a), increasing the liquid local temperature and evaporation rate [158]. There exists an optimum amount of scattering, however, as too high levels can induce backscattering which reduces the amount of energy absorbed in the fluid. Hybrid optoplasmonic nanofluids allow separation of absorbing and scattering centers to achieve greater tunability of the heat localization and phase transformations [254]. Another approach to localize the absorbed light and heat on the air-water interface is through the use of locally high densities of nanoparticles, e.g., by creating thin layers of nanoparticles floating on the surface of the water supported by thermally-insulating substrates [243,255] (Fig. 16b).

Achieving efficient absorption of the broadband solar spectrum is a challenge in SPP-driven solar desalination technologies. Resonance-based plasmonic phenomena typically result in spectrally-localized absorption peaks. In contrast, the sun emits broadband radiation at wavelengths between 350nm-2.5μm. A promising approach to achieve broadband absorption is using multi-scale plasmonic structures. For example, broadband absorption can be achieved using bundles of metallic nanowires arranged in random patterns, creating multiscale features ranging in size from zero to hundreds of nanometers [244]. Another example of a broadband plasmonic absorber with 91% absorptance in the solar spectrum is a multi-scale nanopatterned membrane with embedded nanoparticles of varying size [256].

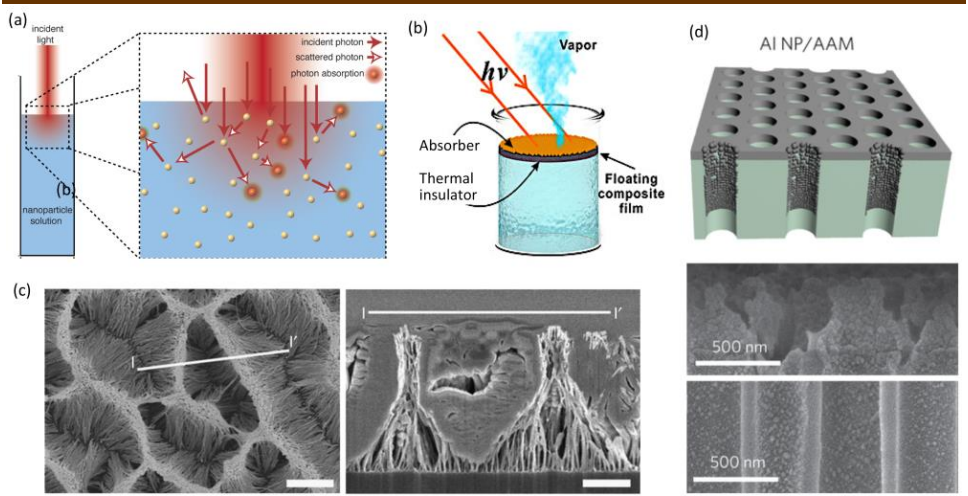

**Fig. 16. Localized plasmonic heating for solar evaporation and water desalination.** (a) Strong SPP-mediated scattering by Au nanoparticles dispersed in water localizes light and heat on the interface between water and air. Reproduced with permission from [158], ©ACS. (b) Floating double-layered film (top: light-to-heat conversion Au nanoparticle layer; bottom: supporting thermally insulating layer). Reproduced with permission from [255], ©NPG. (c) Mesoscale Au nanowire membranes for broadband light absorption. Reproduced with permission from [244], ©NPG. (d) Mesoscale photonic crystal membranes incorporating nanoparticle assemblies in the pores. Reproduced with permission from [256], ©NPG.

## 7. THERMAL EXCITATION OF SURFACE PLASMONS

By the virtue of Kirchhoff's law of radiation, the spectral directional absorbtance is equal to the spectral directional emittance [197,257]. Therefore, the design of SPP-mediated absorbers can be naturally extended to the design of thermal emitters. SPP modes with high local density of photon states can strongly enhance and shape thermal emission both spectrally and angularly. Below, we discuss some applications of thermally-activated SPP modes in both far- and near-field radiative heat transfer.

## 7.1 Surface plasmon-enhanced partially-coherent thermal emission

Thermal emission is a spontaneous process and thus is typically incoherent and unpolarized. According to the Planck's law of blackbody radiation, the intensity of isotropic thermal radiation from the material surface at temperature $T$ per unit projected area, per unit solid angle, and per unit frequency $\omega$ is $U(\omega) = \sigma(\omega) \cdot \left( \hbar\omega^3 / 4\pi^3 c^2 \right) \cdot \left( \exp\{\hbar\omega/k_B T - 1\} \right)^{-1}$, where $k_B$ is the Boltzmann constant and $\sigma(\omega)$ is the graybody absorption rate [258,259]. This well-known formula assumes the free-space photon density of states $D(\omega) = \omega^2/\pi^2 c^3$ and the photon group velocity equal to the speed of light $v_g = c$. However, just like the crystal lattice shapes the electron bandstructure and modifies the Fermi surface and the electron density of states (see Fig. 6), nanostructured plasmonic surfaces can significantly modify both photon DOS and group velocity from their free space values [27,197]. Since intensity of radiation is directly proportional to both $v_g(\omega)$ and $D(\omega)$, spectral characteristics of the emitter can be sculpted by either suppressing or enhancing radiation intensity at a given frequency or into a given direction.

Resonant effects associated with the excitation of surface polariton waves are being actively explored to develop thermal emitters with (partial) spectral and spatial coherence [197,260–262]. As a simple example of nanostructured thermal emitters, periodic gratings made of materials supporting either SPP or SPhP modes (Fig. 17a) exhibit strong narrow resonances in their thermal emittance spectra (Fig. 17b) [263].

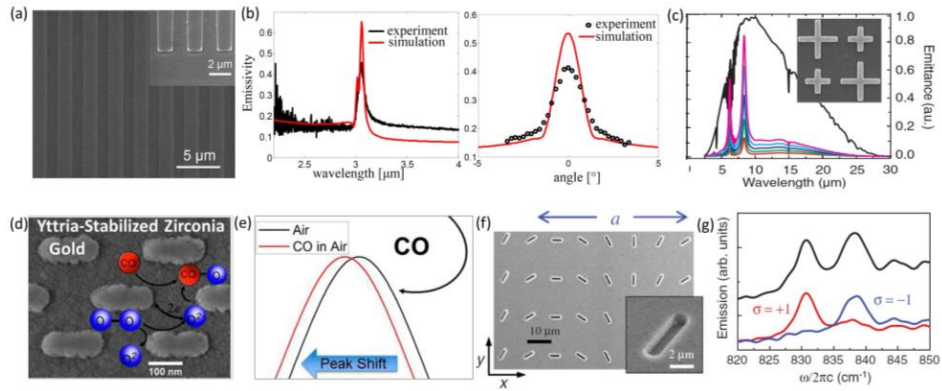

**Fig. 17. Thermally activated SPP modes yield spectrally-selective and directional thermal photon sources.** (a) SEM of a TiN grating that exhibits SPP-mediated coherent thermal emission. (b) Frequency and angular spectra of thermal emission from the plasmonic grating. Panels (a,b) are reproduced with permission from [263], ©OSA. (c) Dual-wavelength Au metamaterial IR emitter. Reproduced with permission from [264], ©APS. (d) SEM of a plasmonic emitter-enabled gas sensor. (e) Thermal emission spectrum of plasmonic sensor shifts in the presence of target gas species, enabling their detection. Panels (d,e) are reproduced with permission from [265], ©ACS. (f) SEM of an array of nanorod plasmonic emitters with spatially-varying local anisotropy axis. (g) Circularly-polarized thermal emission spectra with two distinct central wavelengths exhibited by a plasmonic array shown in panel f. Panels (f,g) are reproduced with permission from [266], ©APS.

Various configurations of narrow-bandwidth thermal emitters have been demonstrated, including those based on perforated metal membranes [267], 'bull's eye' grating structures [268], microstrip patches [269], plasmonic gap nanoantennas [270], periodic antenna arrays [271,272], plasmonic metamaterials and metasurfaces [273–275].

**Selective thermal emitters** based on excitation of SPhP modes in polar dielectrics and those making use of luminescent bands of rare earth oxides are limited by the availability of materials, while plasmonic nanostructures allow wide spectral tunability and multi-band emission spectra. Dual-and multi-band thermal emission has been demonstrated to date with a variety of plasmonic nanostructures and metamaterials [264,276–278]. Tunable frequency-selective narrowband plasmonic emitters can provide currently missing infrared and terahertz light sources for sensing and spectroscopy [279] as well as selective photon sources for thermophotovoltaic (TPV) energy conversion platforms [155,280–282]. In particular, gas sensing applications benefit significantly from the development of thermal plasmonic emitters, as they eliminate the need for external light sources. It has been shown that detection of gases such as $H_2$, CO, and $NO_2$ in combustion environments at temperatures above 500°C or greater can make use of thermally-activated SPP-mediated thermal emission for reliable detection instead of using a laser source (Figs. 17 d,e) [265,283,284]. This makes it possible to reduce gas sensor cost, improve reliability as well as simplifying sensor design and integration into industrial setups.

Finally, thermal emission from nanoscale sources can be tuned to exhibit tailored polarization characteristics [285]. Nanowires and nanorods made of materials supporting SPP and SPhP waves emit photons that are strongly polarized in the direction orthogonal to the wire [286,287]. By arranging polarization-anisotropic emitters into patterned metasurfaces, surface thermal emission can be shaped further. For example, spin-symmetry breaking in thermal radiation can be achieved by introducing local anisotropy in the alignment of nanorod plasmonic antennas on a planar substrate (Fig. 17 f,g) [260,288,266].

### 7.2 Surface plasmon-enhanced near-field radiative heat transfer

When two objects are separated by distances comparable to the wavelength of thermal radiation, radiative exchange can also occur by evanescent tunneling of radiative modes confined within or near the boundaries of the emitting medium. In this so-called **near-field regime**, radiative heat transfer can greatly exceed the far-field blackbody limit as evanescent tunneling enables a greater number of radiative modes to participate in energy exchange [197,262,289–296]. Generally, there are two primary sources of evanescent modes for near-field radiative heat transfer: total internal reflection and surface polariton modes (Fig. 18a). The dispersion relation for an isotropic bulk material with refractive index $n$ has the well-known form $|\mathbf{k}| = n\omega/c$. Accordingly, the number of modes contributing to the heat transfer increases over the far-field case (Fig. 18a) [262]. The photon DOS of an isotropic bulk material is enhanced over that in vacuum and takes the value $D_n(\omega) = n^3 \omega^2 / \pi^2 c^3$. The corresponding radiative intensity increases by the factor of $n^2$ rather than $n^3$ after accounting for the speed of light change in bulk material [197,297,298].

In turn, excitation of surface polariton modes yields a large number of radiative modes within a narrow frequency range (Fig. 18a). The number of participating modes with progressively larger momenta increases with the narrowing gap size $d$, resulting in a substantial enhancement to the DOS and thus to near-field radiative heat transfer. In reality, the cut-off momentum value is smaller than $1/d$, and limited by the dissipative losses in the material (see Fig. 2e).

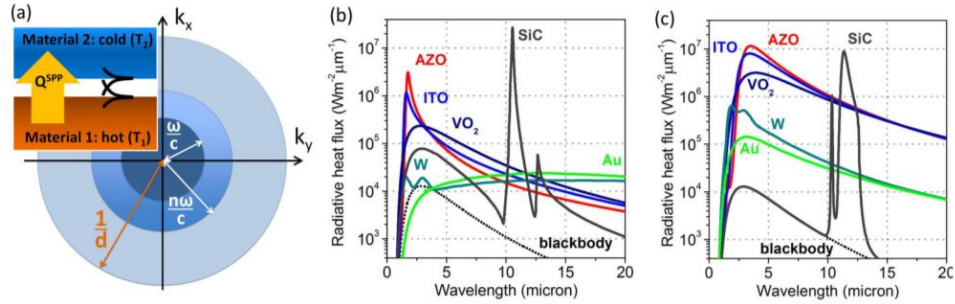

**Fig. 18. Thermally-excited SPP modes enhance near-field radiative heat transfer by orders of magnitude.** (a) Contributions of small- and large-momentum photons to the radiative heat transfer across narrow gaps between an emitter and an absorber. The inset shows a schematic of the radiative heat exchange between hot and cold parallel plates and highlights the contribution of the surface modes to this process. (b) Radiative heat flux across a 20nm gap between surfaces at 300K and 1000K temperature supporting SPhP (SiC) and SPP modes as the function of wavelength. (c) Same as (b) but for heat exchange between 2nm-thin films of the same materials. Panels (b,c) are reproduced with permission from [89], ©MDPI.

The majority of near-field heat transfer studies to date have focused on the use of SPhP modes, which, for most polar dielectrics, overlap spectrally with the Planck distribution near room temperature (see Figs. 7b,c). As a result, all radiative modes associated with the SPhP resonance can be excited thermally with relative ease. The spectral range of SPhP modes has enabled experimental studies observing near-field enhancement at submicron gap separations [96,299–301]. In contrast, SPP modes typically occur at much shorter wavelengths (see Fig. 7a,c), which require very high temperatures to thermally excite. It is for this reason that earlier studies observed limited near-field enhancement between closely spaced metals, which support SPP modes in the UV wavelength range [302–304]. With the advent of new plasmonic materials (see Fig. 7) SPP modes can be excited from the near-IR to the THz wavelength range. Dissipative losses of these plasmonic materials are typically larger than those in polar dielectrics, resulting in damping of the near-field heat transfer (Fig. 18b). However, it was shown recently that SPP modes supported in ultra-thin films of plasmonic materials exhibit significant spectral broadening due to the hybridization of SPP modes on both sides of the film [89]. This spectral broadening in turn increases the total heat flux exchanged by the plasmonic films, increasing it by over an order of magnitude (Fig. 18c).

Although the practical implementation of near-field radiative heat transfer remains very much in its infancy, numerous applications have nonetheless emerged that utilize near-field enhancement associated with SPP modes. The most immediate application of plasmonic near-field enhancement is heat-assisted magnetic recording (see Fig. 12d) [183,185]. Thermophotovoltaics is another promising application, which can benefit from plasmonic near-field enhancement. Theoretical studies have predicted near-field radiative heat transfer can significantly improve the overall performance of TPV platforms [273,305–307].

## 8. SUMMARY AND OUTLOOK

Plasmonic losses stem from the nature of the plasmon-polariton mode's excitation and thus cannot be completely eliminated. Losses can however be reduced by the choice of material system, reduction of temperature, surface polishing, and judicious choice of the nanostructure geometry. In particular, hybrid nanostructures that separate light trapping and SPP-excitation sites offer significant advantages in narrowing and tuning of SPP modes. The unique nature of plasmonic excitation and decay translates into new regimes of light-induced charge carrier generation and harvesting, which are not possible under the direct excitation by small-momentum propagating photons. Hybrid integration extends the use of conventional plasmonic materials such as noble metals into the IR spectral domain, which is otherwise not easily accessible. New plasmonic materials continue to emerge, promising low-loss plasmonic performance in spectral ranges from the visible to the THz. However, some applications such as those in the energy generation are quite sensitive to the material cost factor, which will shape the new materials discovery strategy as well as strategies to recycle precious materials such as noble metals. Many applications, e.g. catalysis and medicine, will continue to strongly benefit from a combination of functionalities provided by localized SPP modes, including light and heat localization, which in turn can provide remote temperature readout mechanism. Technological advances in nanofabrication are expected to bring about the era of using the SPP-enhanced near-field radiative energy transfer, which can translate into significant improvements in the next-generation energy conversion devices.

**Funding.** This work was supported by DOE-BES Award No. DE-FG02-02ER45977 (for thermal emission tailoring and near-field radiative heat transport), the ARPA-E award DE-AR0000471 (for full solar spectrum harvesting), and the 'Solid State Solar-Thermal Energy Conversion Center (S3TEC)', funded by the US Department of Energy, Office of Science, and Office of Basic Energy, Award No. DE-SC0001299/DE-FG02-09ER46577 (for thermophotovoltaic applications).